%% file: main.tex
\documentclass[manuscript,screen,acmsmall,authorversion,nonacm]{acmart}

\AtBeginDocument{%
  \providecommand\BibTeX{{%
    \normalfont B\kern-0.5em{\scshape i\kern-0.25em b}\kern-0.8em\TeX}}}

\input{tool.tex}

\begin{document}

\title{KADEL: Knowledge-Aware Denoising Learning for Commit Message Generation}

\author{Wei Tao}
\email{wtao18@fudan.edu.cn}
\orcid{0000-0002-1800-1904}
\affiliation{%
  \department{Shanghai Engineering Research Center of AI and Robotics, Academy for Engineering and Technology}
  \institution{Fudan University}
  \city{Shanghai}
  \country{China}}

\author{Yucheng Zhou}
\email{yucheng.zhou@connect.um.edu.mo}
\orcid{0009-0006-9883-5621}
\affiliation{
  \department{State Key Laboratory of Internet of Things for Smart City, Department of Computer and Information Science}
  \institution{University of Macau}
  \city{Macau}
  \country{China}}

\author{Yanlin Wang}
\email{wangylin36@mail.sysu.edu.cn}
\orcid{0000-0001-7761-7269}
\affiliation{
  \department{School of Software Engineering}
  \institution{Sun Yat-sen University}
  \city{Zhuhai}
  \state{Guangdong}
  \country{China}
  \postcode{519082}}

\author{Hongyu Zhang}
\email{hyzhang@cqu.edu.cn}
\orcid{0000-0002-3063-9425}
\affiliation{
  \department{School of Big Data and Software Engineering}
  \institution{Chongqing University}
  \city{Chongqing}
  \country{China}}

\author{Haofen Wang}
\email{carter.whfcarter@gmail.com}
\orcid{0000-0003-3018-3824}
\affiliation{%
  \department{College of Design and Innovation}
  \institution{Tongji University}
  \city{Shanghai}
  \country{China}}

\author{Wenqiang Zhang}
\email{wqzhang@fudan.edu.cn}
\orcid{0000-0002-3339-8751}
\affiliation{
  \department{Engineering Research Center of AI and Robotics, Ministry of Education, Academy for Engineering and Technology; Shanghai Key Lab of Intelligent Information Processing, School of Computer Science}
  \institution{Fudan University}
  \streetaddress{220 Handan Road}
  \city{Shanghai}
  \country{China}
  \postcode{200433}}

\begin{abstract}
Commit messages are natural language descriptions of code changes, which are important for software evolution such as code understanding and maintenance. 
However, previous methods are trained on the entire dataset without considering the fact that a portion of commit messages adhere to good practice (i.e., good-practice commits), while the rest do not.
On the basis of our empirical study, we discover that training on good-practice commits significantly contributes to the commit message generation.
Motivated by this finding, we propose a novel knowledge-aware denoising learning method called \Ourmethod. 
Considering that good-practice commits constitute only a small proportion of the dataset, we align the remaining training samples with these good-practice commits.
To achieve this, we propose a model that learns the commit knowledge by training on good-practice commits. 
This knowledge model enables supplementing more information for training samples that do not conform to good practice.
However, since the supplementary information may contain noise or prediction errors, we propose a dynamic denoising training method. This method composes a distribution-aware confidence function and a dynamic distribution list, which enhances the effectiveness of the training process.
Experimental results on the whole \mcmd dataset demonstrate that our method overall achieves state-of-the-art performance compared with previous methods.
\end{abstract}



\thanks{This work was supported by National Natural Science Foundation of China (No.62072112), Scientific and Technological innovation action plan of Shanghai Science and Technology Committee (No.22511102202).}

\maketitle

\section{Introduction}

A large amount of code is frequently updated, and the code changes drive software development. To better manage software evolution, version control systems such as Git require developers to describe the changes in natural language (i.e., commit messages) each time they update the code. 
Commit messages enable developers to better understand, manage, and analyze software evolution~\citep{DBLP:conf/icse/LiA23}. 
For example, they provide additional explanatory power in code reviewer recommendation~\citep{YeZAM21}, commit classification~\citep{CasalnuovoSRR17}, maintenance activity classification~\citep{HindleGGH09}, refactoring recommendation~\citep{RebaiKASK20}, and just-in-time defect prediction~\citep{BarnettGSM16}.

\begin{figure}[h]
    \centering
    \includegraphics[width=0.8\linewidth]{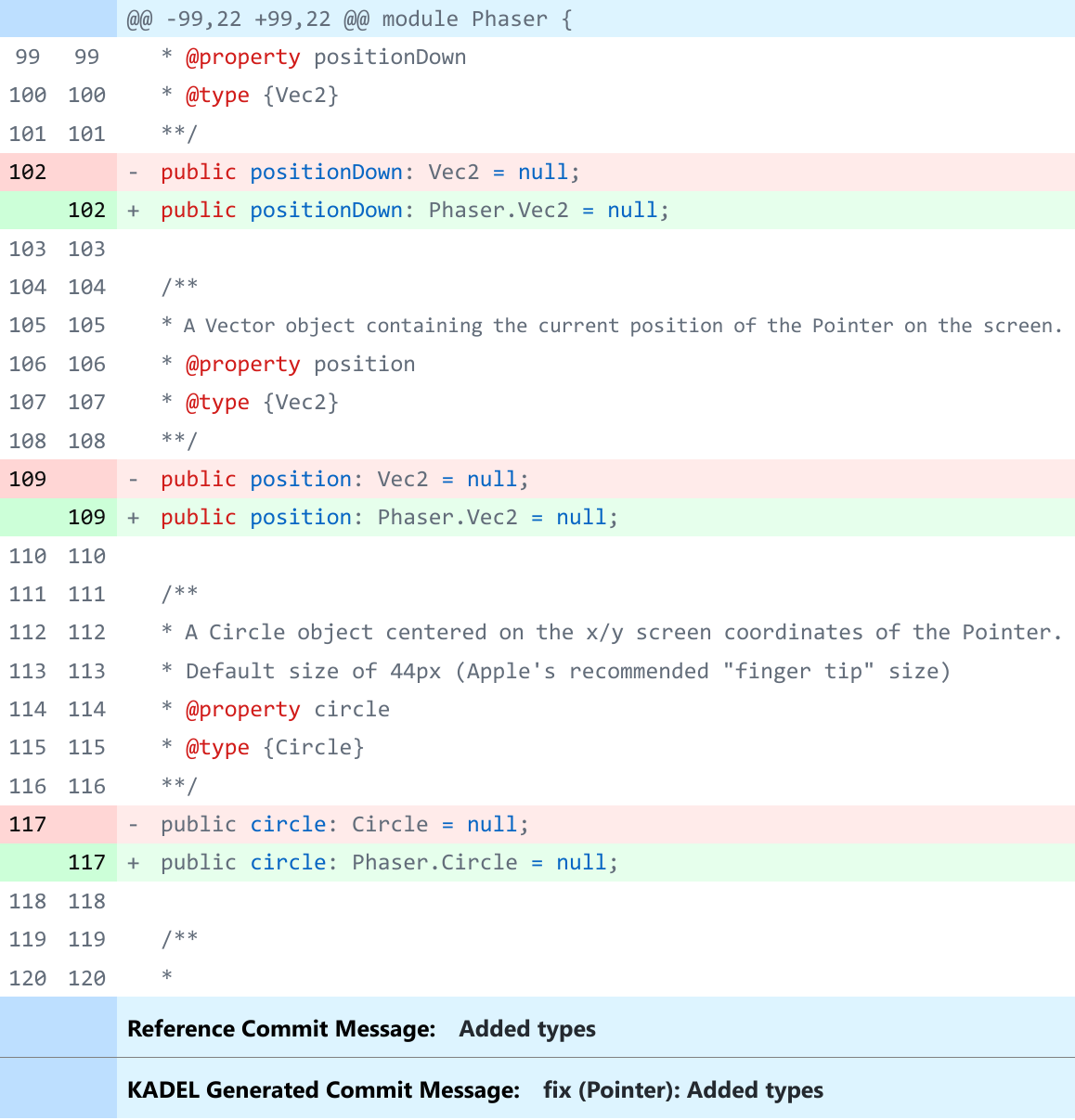}
    \caption{An example of code change and the corresponding commit message from GitHub.}
    \label{fig:example_commit}
    \Description{.}
\end{figure}

Over the past years, a number of approaches have been proposed to generate commit messages automatically. 
Early works explored rule-based methods, utilizing predefined rules to generate commit messages~\citep{Cortes-CoyVAP14,VasquezCAP15,BuseW10}.
Subsequently, information retrieval based techniques were also introduced to overcome the constraint of predefined rules~\citep{LiuXHLXW18,HuangJZCZT20}.
Recently, various deep learning-based models have been proposed for commit message generation. 
These studies~\citep{LoyolaMBMS18,Jiang19,LiuLZFDQ19,Xu00GT019,Liu20Atom,DongLZ0LZ022,ShiW0DZHZ022} significantly improved the quality of the generated commit messages, demonstrating the potential of deep models in code change understanding.

Despite their success, these methods are trained on the entire dataset without considering the fact that a portion of commit messages adhere to good practice while the rest are poor training samples.
For instance, as shown in \Fig~\ref{fig:example_commit}, the reference commit message ``Added types'' fails to provide a clear explanation of why the developer makes this commit\footnote{\url{https://github.com/photonstorm/phaser/commit/c2f0128}} and what the scope of this code change is. 
According to the standard~\citep{TianZSJL22}, this is not considered a good commit message. 
Such poor training samples abound.
\citet{TianZSJL22} studied five open-source software and found that 44\% of the commit messages are of poor quality. 
The reason is that insufficient time, experience, or willingness makes the quality of commit messages varies.
To write informative and easily understandable commit messages, some projects, such as AngularJS, defined and followed good practice, ``precise rules over how our git commit messages can be formatted''\footnote{Angular Commit Guidelines: \url{https://github.com/angular/angular.js/blob/master/DEVELOPERS.md/\#commit-message-format}}.
This AngularJS commit rule is often mentioned as a good practice in many repositories' commit guidelines, especially in JavaScript communities.
Following this good practice, the commit message shown in \Fig~\ref{fig:example_commit} could be improved by prepending it with ``fix (Pointer)'', where ``fix'' and ``Pointer'' represent the type and scope of the commit, respectively.
This better commit message requires the developers to use their knowledge of development
and maintenance to follow this rule and generate this information (i.e., type and scope).
Such information helps developers understand that this commit is for bug-$fix$ing and the scope of this code change is about the object, $Pointer$, which is also mentioned in the context comments.
However, we discovered that the majority of commit messages lack this information, which could reduce the clarity of the commit messages.
For example, less than 2.28\% of commit messages in one of the largest publicly available commit message datasets, \mcmd~\citep{TaoWSDH0ZZ21}, contain this specific information.

In this paper, first, we empirically studied how the good practice (AngularJS rule) affects the automatic commit message generation. We discover that training on the examples following the good practice can significantly contribute to the commit message generation.
Inspired by the above finding, we propose a \textbf{K}nowledge-\textbf{A}ware \textbf{DE}noising \textbf{L}earning (\textbf{\Ourmethod}) method to introduce commit knowledge to the model hence improving the training data and enhancing the effectiveness of training process. 
Motivated by previous studies \citep{BosselutRSMCC19,ZhouGSPZJ21} that train a knowledge model as a provider of commit knowledge, we propose a model trained on data following the good practice.
Taking each pair of code changes and commit messages as input, the commit knowledge model learns to predict the type and scope of a commit message. 
After being trained, the knowledge model can enrich the original commit messages with the type and scope of code changes.

Although the commit knowledge model can provide type and scope information, 
it is inevitable that noise (i.e., error prediction) exists. 
To learn with label noise, numerous studies have proposed various techniques to improve model learning by loss correction~\citep{PatriniRMNQ17}, loss reweighting~\citep{LiuT16}, or label refurbishment~\citep{Zhou22Ranker}, etc.
However, these approaches are mainly designed for discriminative tasks, where label probabilities can be used to derive label confidence. 
In contrast to discriminative tasks, generation tasks fail to directly infer confidence of the generated results through the product of label probabilities,  which heavily relies on the length of generated results.
Inspired by the study~\citep{ArazoOAOM19} that demonstrates loss of clean and noisy data following their respective distributions, we leverage the expectation–maximization (EM) algorithm \citep{dempster1977maximum} to deduce two distributions of clean and noise data in training loss.
In addition, we propose a novel dynamic denoising training method that incorporates a distribution-aware confidence function and a dynamic distribution list. 
During model training, we build and update the dynamic distribution list to record training loss and deduce two distributions by the EM algorithm and reformalize training loss by the distribution-aware confidence function based on two distributions. 

In experiments, we train our model on the whole \mcmd dataset and conduct an evaluation on each programming language test set of \mcmd, including rule-unmatched subset (\mcmdplu, which did not follow the good practice) and rule-matched subset (\mcmdplm, which follow the good practice). Experimental results show that our method overall achieves state-of-the-art performance compared with other strong competitors. Moreover, we investigate the effectiveness of our method through extensive analysis including human evaluation.

Contributions of this work are summarized as follows:
\begin{itemize}
    \item We empirically study commit messages in \mcmd and find that commit knowledge can be extracted from the commits following the good practice \revise{and it contributes to commit message generation}.
    \item We propose a novel method, \Ourmethod, for commit message generation. In the method, we build a commit knowledge model trained on data following good practice and design a novel dynamic denoising training method that composes a distribution-aware confidence function and a dynamic distribution list to achieve more effective training.
    \item Experimental results show that \Ourmethod overall achieves the state-of-the-art performance in commit message generation and each component in \Ourmethod is effective.
\end{itemize}

\section{Related Work}

\subsection{Commit Message Generation}

Over the past years, many approaches have been proposed to generate commit messages automatically.

Early work~\citep{VasquezCAP15,BuseW10,Cortes-CoyVAP14} is based on expert rules. However, these rule-based methods tend to generate long commit messages with too many lines, which are difficult to convey the key intention of the code changes. 

Later, information retrieval based techniques are introduced to commit message generation~\citep{LiuXHLXW18,HuangJZCZT20}. For instance, \citet{LiuXHLXW18} propose a simple yet effective retrieval-based method utilizing the nearest neighbor algorithm.

\citet{HuangJZCZT20} propose to retrieve the most similar commits according to the syntax and semantics in the changed code.

Recently, deep learning-based techniques are utilized for commit message generation. Some studies~\citep{JiangAM17,LoyolaMBMS18,LoyolaMM17,Jiang19} represent code changes as textual sequences and use NMT techniques to translate the source code changes into target commit messages. 

\citet{LiuLZFDQ19} adopt the pointer-generator network~\citep{SeeLM17} to handle the out-of-vocabulary problem. Some studies leverage the rich structural information of source code. \citet{Xu00GT019} attempted to model both the semantic representation and structural representation of code changes, \citet{Liu20Atom} capture the abstract syntax tree structure of code changes and its semantics. \citet{DongLZ0LZ022} represent code changes as fine-grained graphs. This structural information helps models learn to generate commit messages automatically. \citet{ShiW0DZHZ022} proposed \Race which combines information retrieval techniques with learning-based generation methods. \revise{\citet{HeWWZZ023} proposed COME which combines retrieval techniques with translation-based methods through a decision algorithm and this method learns better contextualized code change representation.}

Although these methods show great performance, the good practice that condenses the wisdom of the development and maintenance community is ignored. For example, the AngularJS rule is often mentioned in many repositories' commit guidelines, especially in JavaScript communities. In this paper, we want to take full advantage of this rule and introduce the commit knowledge to our model hence generating better commit messages.

\subsection{Knowledge-Augmented Language Models}

Pre-trained language models (PLMs) have made remarkable progress in recent years, and they have demonstrated their effectiveness for various text and code tasks by fine-tuning them \citep{LvZH20,ZhouGSPZJ21}.

However, PLMs still encounter a challenge, namely that they have limited memory capacity and knowledge.

With the advance in large-scale knowledge graphs \citep{SapBABLRRSC19}, an effective retrieval-based paradigm incorporates knowledge into language models to improve their reasoning or generation capability \citep{LvZH20,Wang22Unsupervised}.

\citet{LvZH20} employ BM25 to retrieve relevant knowledge from external knowledge graphs based on the natural language input, which enriches the model’s prior understanding and enhances its reasoning performance.

\citet{Chen19Enhancing} propose a method to enhance data-to-text generation models with external knowledge to improve the accuracy and informativeness of the generated texts.

Although this paradigm is proven effective, it heavily relies on large-scale knowledge graphs to circumvent the coverage problem (i.e., failure to retrieve relevant knowledge).

Since labeling large-scale knowledge graphs are expensive, \citet{ZhouGSPZJ21} propose to learn a modeling-based knowledge model, which can introduce knowledge for unlabeled examples.

Therefore, we follow this paradigm to inject commit knowledge for commit message generation.

\subsection{Learning with Noisy Labels}

Learning with noisy labels is an essential problem in weakly supervised learning, which aims to improve the generalization ability of models in the noisy labels \citep{MiyatoMKI19}. 

Some methods select samples with small loss values for model training \citep{HanYYNXHTS18,WeiFC020}. For instance, co-teaching \citep{HanYYNXHTS18} is a method that trains two networks simultaneously and allows them to exchange feedback with each other using the selected samples. Therefore, they can reduce the influence of noisy labels and enhance their generalization performance.

Nevertheless, there is a potential pitfall in selecting samples with small loss values as they may fail to accurately represent the true data distribution, resulting in overfitting. To circumvent this predicament, some methods \citep{PatriniRMNQ17,HendrycksMWG18} advocate for the employment of an estimated noise transition matrix for the purpose of loss correction and the adaptive allocation of weights to samples throughout the training phase.

Furthermore, a popular strategy to enhance the performance of learning models is to assign higher weights to clean samples \citep{LiuT16,JiangZLLF18}.

In this work, we predict two distributions of clean and noisy data and design a distribution-aware confidence function to re-weight samples.

\section{Empirical Study}\label{sec:empirical}

In this section, we want to investigate what is the good practice in creating the commit, whether the good practice (AngularJS commit rule) influences commit message generation and how can we use the good practice to improve the generation. \revise{Furthermore, we analyze the empirical findings and discuss the possible reasons for them.}

\subsection{\revise{Experiments and Empirical Findings}}

As \mcmd has good traceability to find each commit's source, we investigated all of the repositories in \mcmd ~\revise{by manually inspecting the file contents of each repository root directory and associated documentation such as README.md} and find that 188 out of 500 repositories do not have contributor guidelines\footnote{\revise{Statistics as of October 4, 2022}} which are recommended in GitHub\footnote{\url{https://docs.github.com/en/communities/setting-up-your-project-for-healthy-contributions/setting-guidelines-for-repository-contributors}}. 

Although other repositories have contributor guidelines, only a few of them have clear rules about the commit messages. Specifically, repositories represented by AngularJS\footnote{\url{https://github.com/angular/angular.js}} have precise rules for commit messages: it should ``include a type, a scope and a subject''. This rule is mentioned in contributor guidelines of many JavaScript repositories, which means it is popular as a good practice in these JavaScript communities. Moreover, the rule is also introduced in some developing tools such as Commitizen\footnote{\url{http://commitizen.github.io/cz-cli/}} which is a ``release management tool designed for teams''.

Following the AngularJS rule makes commit messages more meaningful and readable.
For example, as shown in \Fig~\ref{fig:example_commit}, the commit message ``Added types'' tells us what code changes but does not provide the reason why these code changes are made. Another commit message, ``fix (Pointer): Added types'', not only provide what changes but also explain the reason, this commit is used for bug-fixing and the scope of the change is the Pointer (the context code explains that Phaser is ``a Pointer object is used by the Touch and MSPoint managers and represents a single finger on the touch screen.''). The latter message can be regarded as a good commit message according to the standard~\citep{TianZSJL22}.

Although the convention rule is welcome in many JavaScript repositories, the number of commits that matched the rule is still relatively limited. For each commit message in \mcmd, we use a regular expression to determine whether it matches the AngularJS rule. The corresponding statistics of \mcmd are shown in \Tab~\ref{tab:mcmd_stats}.

As shown in \Tab~\ref{tab:mcmd_stats}, there are 8.70\% commit messages in \mcmdjs matching the AngularJS rule while less than 1\% commit messages in other programming languages' repositories match. This difference means that the popularity of the rule in JavaScript communities is significantly higher than in others.

One possible reason is that providing ``type'' and ``scope'' is hard and time-consuming, especially for new developers. Understanding the meaning of each ``type'' and classifying the commit into the right ``type'' requires the developers to have the knowledge of development and maintenance.
It is important to investigate whether the rules influence the developers in writing commit messages and what are the impacts on the generation model. We take the pre-trained programming language model \Codetfive~\citep{0034WJH21} as an example to analyze the impact of commit knowledge on the message generation. \Codetfive is chosen because it shows state-of-the-art performance on generation tasks in the benchmark CodeXGLUE~\citep{LuGRHSBCDJTLZSZ21}\footnote{\url{https://microsoft.github.io/CodeXGLUE/}}.

\begin{table}[!t]

\centering
\caption{The statistics of rule-matched commit messages in the \mcmd.}

\label{tab:mcmd_stats}

\begin{tabular}{lcc}

\toprule
\textbf{Data}   & \textbf{\# Matching Messages}       & \textbf{Ratio}\\

\midrule
\mcmdjs         & 39165     & 8.70\%    \\
\mcmdcsharp     & 3705      & 0.82\%    \\
\mcmdpython     & 3431      & 0.76\%    \\
\mcmdcpp        & 3148      & 0.70\%    \\
\mcmdjava       & 1799      & 0.40\%    \\
\toprule
\end{tabular}

\end{table}

According to the definition, a rule-matched commit message can be split into three components: type, scope, and subject. 

Real rule-matched commits are suitable to do experiments and \mcmdjs has the largest number of rule-matched examples so we select the commits in which the corresponding message matches the rule from \mcmdjs, which is denoted as \mcmdjsm.

Based on the \mcmdjsm, we fine-tuned \Codetfive with different content components.

\revise{As the length of the type is one and the length of scope is short (those with no more than 3 words account for more than 93\%), evaluating the performance of the generation of them can be considered as a multi-label classification task. Therefore, EM and }F1-score are used to evaluate the performance of type and scope. 

We use \bleunorm, which is demonstrated as a good BLEU variant~\citep{TaoWSDH0ZZ21}, to evaluate the performance of the subject.

Some research works~\citep{BulteT19, LewisPPPKGKLYR020} have explored introducing the knowledge to the encoder of the pre-trained model. 

Therefore, we straightforwardly cooperate the ``type'' and(/or) ``scope'' with code changes in the encoder input. This training setting is to simulate the experts to give the ``type'' and/or ``scope'' information with code changes to the model in the input and the model can generate the ``subject'' of the commit message based on them.

As shown in \Tab~\ref{tab:codetfive-mcmdjs-different-input}, the experimental results in this setting demonstrate that different settings of encoding input do not influence the performance of the subject in general. It indicates that commit knowledge is hard to be introduced to the model by adding the type and scope in the encoder.

\begin{table}
\centering
\caption{The performance of \Codetfive fine-tuned on \mcmdjsm in different settings of the encoder input.}
\label{tab:codetfive-mcmdjs-different-input}
\begin{tabular}{ccccc}
\toprule
\multicolumn{4}{c}{\textbf{Training Setting}}  & \multicolumn{1}{c}{\textbf{Test Performance}} \\
\cmidrule(r){1-4} \cmidrule(r){5-5}
\multicolumn{3}{c}{Encoder Input} & Decoder Output  & \revise{Subject} \\
\cmidrule(r){1-3} \cmidrule(r){4-4} \cmidrule(r){5-5} 
 Type & Scope & \cdiff &\subject  & \bleunorm  \\
\cmidrule(r){1-3} \cmidrule(r){4-4} \cmidrule(r){5-5} 
$\checkmark$ & $\checkmark$ & $\checkmark$  & $\checkmark$  & 22.30   \\
$\checkmark$ &  & $\checkmark$   & $\checkmark$     & 22.35  \\
 & $\checkmark$  & $\checkmark$  & $\checkmark$     & 22.02  \\
 &   &  $\checkmark$  & $\checkmark$        & 22.34  \\
\toprule
\end{tabular}
\end{table}

Attaching ``type'' and ``scope'' with code changes in the encoder does not work but the AngularJS rule is helpful when human developers write the commit message (as described in the second paragraph in this section). How can we effectively introduce the commit knowledge into the model based on the rule-matched commits?

As the saying goes, ``Give a man a fish, and you feed him for a day. Teach a man to fish, and you feed him for a lifetime.'', attaching the external information (``type'' and ``scope'') in the encoder is like giving the model a ``fish'' but we want the model to have the ability to ``fish'', which means it can generate ``type, scope'' with commit knowledge. 

Inspired by this, we put the ``type'' and(/or) ``scope'' with ``subject'' in the decoder output during training. It simulates the process of learning to generate the components of ``type'' and(/or) ``scope'' and the model can generate content of all components after training.
As shown in \Tab~\ref{tab:codetfive-mcmdjs-different-output},  the experimental results show that \bleunorm of the \Codetfive trained with \type and \scope is higher than the score of \Codetfive trained without \type or \scope, indicating that this setting is helpful to generate better \subject and commit knowledge is probably introduced in the model. 

\begin{table}
\centering
\caption{The performance of \Codetfive fine-tuned on \mcmdjsm in different settings of the decoder output. 
A checkmark means the data item is used for training in that setting.}
\label{tab:codetfive-mcmdjs-different-output}
\begin{tabular}{ccccccccc}
\toprule
\multicolumn{4}{c}{\textbf{Training Setting}}  & \multicolumn{5}{c}{\textbf{Test Performance}} \\
\cmidrule(r){1-4} \cmidrule(r){5-9}
Input & \multicolumn{3}{c}{Decoder Output}    & \multicolumn{2}{c}{\revise{Type}} & \multicolumn{2}{c}{\revise{Scope}} & \revise{Subject} \\
\cmidrule(r){1-1} \cmidrule(r){2-4} \cmidrule(r){5-6} \cmidrule(r){7-8} \cmidrule(r){9-9}

\cdiff & Type & Scope & \subject  & \revise{EM}  & F1    & \revise{EM}    & F1 & \bleumosesnorm  \\

\cmidrule(r){1-1} \cmidrule(r){2-4} \cmidrule(r){5-6} \cmidrule(r){7-8} \cmidrule(r){9-9}

$\checkmark$ & $\checkmark$ & $\checkmark$  & $\checkmark$  & \revise{\textbf{62.27}}    & \textbf{61.59} & \revise{\textbf{57.39}} & \textbf{56.67}  & \textbf{27.70}   \\

$\checkmark$ & $\checkmark$ &    & $\checkmark$     & \revise{56.77}    & 54.42     & \revise{-}  & -  & 22.81  \\
$\checkmark$ &   & $\checkmark$  & $\checkmark$     & \revise{-} & - & \revise{45.32}    & 45.37  & 21.37  \\
$\checkmark$ &   &    & $\checkmark$        & \revise{-} & - & \revise{-} & -  & 22.34  \\
\toprule
\end{tabular}

\end{table}

Compared with the results shown in \Tab~\ref{tab:codetfive-mcmdjs-different-input} and \Tab~\ref{tab:codetfive-mcmdjs-different-output}, we can infer that the latter setting is a better way to introduce commit knowledge into the model and it has more potential to improve the ``subject'' generation.

\subsection{\revise{Analysis and Discussion}}

\revise{The empirical finding raises two critical questions for further exploration: (1) Why is it beneficial for the model when integrating ``type'' and ``scope'' information into the decoder instead of the encoder during training?  (2) Why does the model benefit greater from the combination of two types of information (``type'' or ``scope'') during training?}

\begin{figure*}
    \includegraphics[width=0.32\columnwidth]{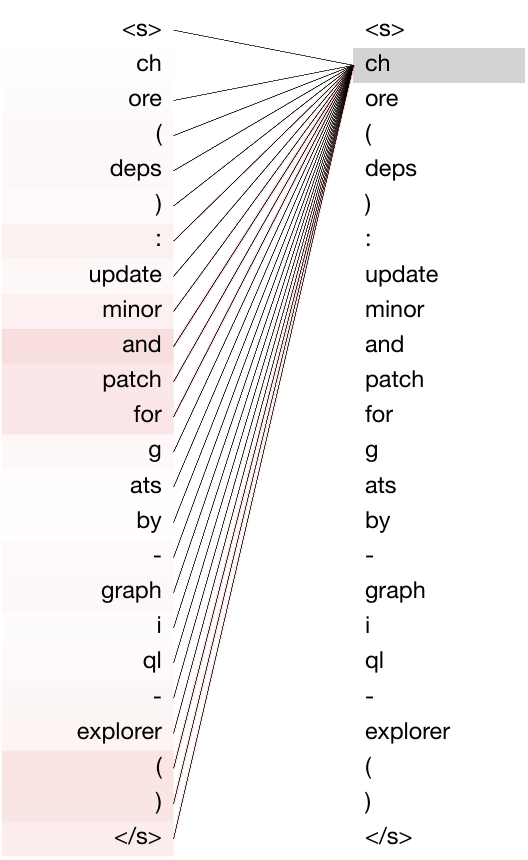}\label{b}
    \includegraphics[width=0.32\columnwidth]{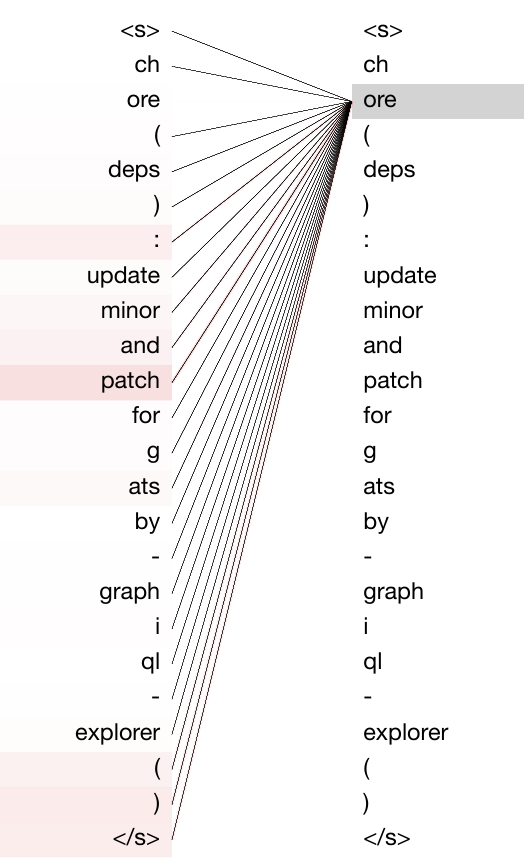}\label{a}
    \includegraphics[width=0.32\columnwidth]{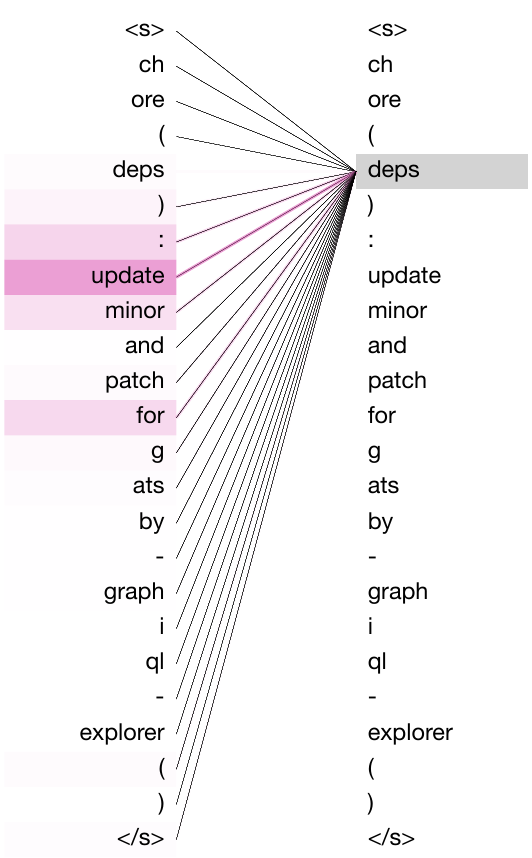}\label{c}
    \includegraphics[width=0.32\columnwidth]{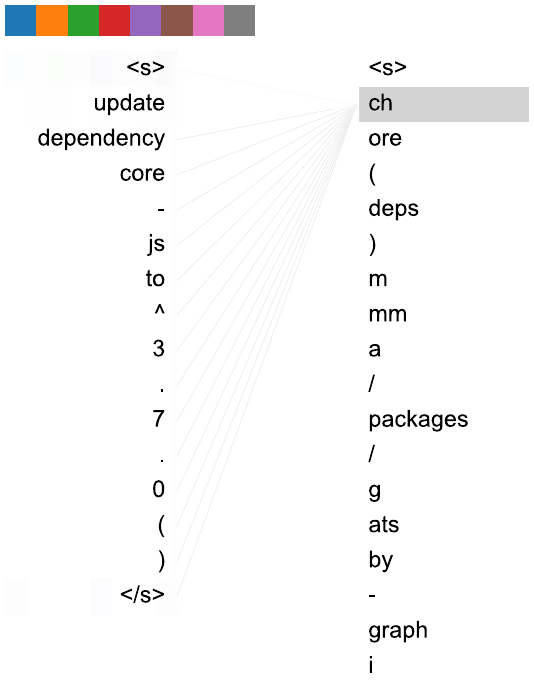}\label{d}
    \includegraphics[width=0.32\columnwidth]{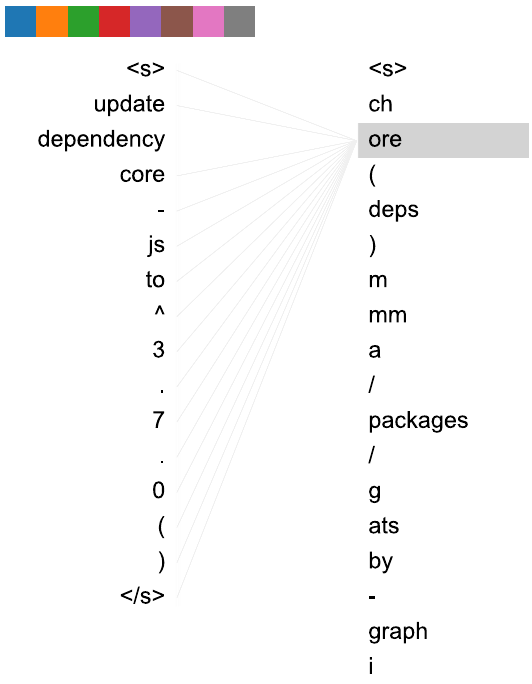}\label{e}
    \includegraphics[width=0.32\columnwidth]{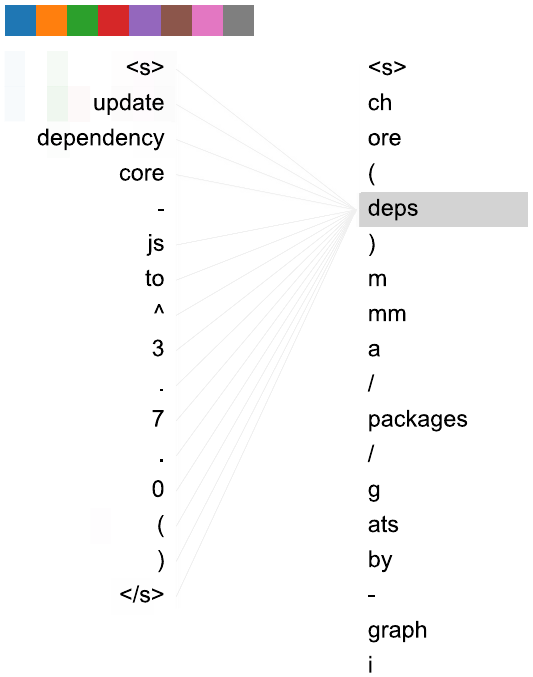}\label{f}
    \caption{\revise{The three sub-figures in the upper row show the decoder attention weights of the model trained on the setting of decoder output containing ``type'' and ``scope''. The lower row shows the cross attention weights of the model trained on the setting of encoder input containing ``type'' and ``scope''. Rectangular blocks of different colors represent different head attention. The darker the color, the higher the attention weight.}}
    \label{fig:attention-weight-enc-dec}
\end{figure*}

\revise{To answer the first question, we conduct an in-depth analysis of the generation process in trained models. Using the visualization tool~\citep{vig-2019-multiscale}, we observe varying model's attention weights on ``type'' and ``scope'' during the generation process. The attention weights of the model trained in different settings are shown in \Fig~\ref{fig:attention-weight-enc-dec}. As the sub-figures in the upper row of \Fig~\ref{fig:attention-weight-enc-dec} show, the model trained with both the ``type'' and ``scope'' information in the decoder can pay attention to both of them during the generation of tokens of the ``subject'' part. Conversely, models with ``type'' and ``scope'' information in the encoder show negligible utilization of these aspects in generating ``subject'' content. This variance in attention allocation contributes to performance disparities in the generation, thereby answering the first research question.}

\begin{table}
\centering
\caption{\revise{The performance of \Codetfive fine-tuned on \mcmdjsm in different settings of the decoder output. The test set is divided into two categories: one category is that the content before ``subject'' is correctly predicted (denoted as Prefix Correct), and the other category is not (denoted as Prefix Wrong).
Diff, T, S, BLEU, \rouge is short for code changes, Type, Scope, \bleunorm, and \rougel respectively. 
A checkmark means the data item is used for training in that setting.
The numbers in brackets represent the difference between the respective scores and the scores in the same test set under the settings of the last row of \Tab~\ref{tab:codetfive-mcmdjs-different-output}.
}}
\label{tab:codetfive-mcmdjs-different-output-subject-split}
\begin{tabular}{cccccccccc}
\toprule
\multicolumn{4}{c}{\textbf{\revise{Training Setting}}}  & \multicolumn{6}{c}{\textbf{\revise{Test Performance on Subject}}} \\
\cmidrule(r){1-4} \cmidrule(r){5-10}
\revise{Input} & \multicolumn{3}{c}{\revise{Decoder Output}}    & \multicolumn{3}{c}{\revise{Prefix Correct}} & \multicolumn{3}{c}{\revise{Prefix Wrong}} \\
\cmidrule(r){1-1} \cmidrule(r){2-4} \cmidrule(r){5-7} \cmidrule(r){8-10}
\revise{Diff} & \revise{T} & \revise{S} &\revise{\subject}  & \revise{BLEU}  & \revise{\meteor}    & \revise{\rouge}    & \revise{BLEU}  & \revise{\meteor}    & \revise{\rouge}  \\
\cmidrule(r){1-1} \cmidrule(r){2-4} \cmidrule(r){5-7} \cmidrule(r){8-10}
\multirow{2}{*}{\revise{$\checkmark$}} & \multirow{2}{*}{\revise{$\checkmark$}} & \multirow{2}{*}{\revise{$\checkmark$}}  & \multirow{2}{*}{\revise{$\checkmark$}}  &  \revise{45.03} &  \revise{46.74} &   \revise{48.98} &  \revise{16.89} &  \revise{23.65} &   \revise{19.28} \\
&&&&  \revise{(+9.90)} &  \revise{(+10.20)} &   \revise{(+10.54)} &  \revise{(+2.48)} &  \revise{(+3.18)} &   \revise{(+2.49)} \\
\cmidrule(r){1-10}
\multirow{2}{*}{\revise{$\checkmark$}} & \multirow{2}{*}{\revise{$\checkmark$}} & \multirow{2}{*}{}   & \multirow{2}{*}{\revise{$\checkmark$}}     &  \revise{29.03} &  \revise{31.35} &   \revise{32.48} &  \revise{14.72} &  \revise{21.84} &   \revise{16.69} \\
&&&&  \revise{(+1.30)} &  \revise{(+1.63)}  &   \revise{(+1.50)}  &  \revise{(-0.61)} &  \revise{(-0.75)} &   \revise{(-0.70)} \\
\cmidrule(r){1-10}
\multirow{2}{*}{\revise{$\checkmark$}} & \multirow{2}{*}{} & \multirow{2}{*}{\revise{$\checkmark$}}  & \multirow{2}{*}{\revise{$\checkmark$}}     &  \revise{31.06} &  \revise{33.43} &   \revise{33.67} &  \revise{13.32} &  \revise{18.03} &   \revise{15.99} \\
&&&&  \revise{(-0.41)} &  \revise{(-1.22)}  &   \revise{(-0.25)}  &  \revise{(-1.50)} &  \revise{(-1.98)} &   \revise{(-1.80)} \\
\toprule
\end{tabular}
\end{table}

\begin{figure*}
    \includegraphics[width=0.4\columnwidth]{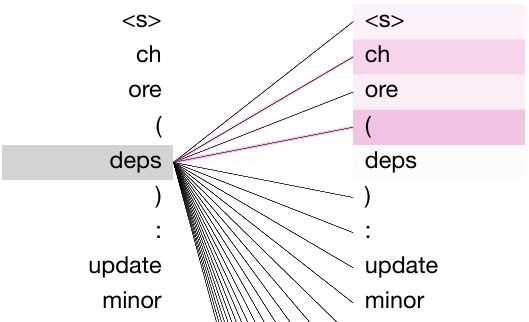}
    \caption{\revise{The decoder attention weights of the model trained on the setting of decoder output containing ``type'' and ``scope''. The darker the color, the higher the attention weight.}}
    \label{fig:attention-weight-scope}
\end{figure*}

\revise{
\Tab~\ref{tab:codetfive-mcmdjs-different-output} shows that the model trained with both ``type'' and ``scope'' information outperforms others in the generation of these two kinds of information. Therefore, a possible reason for the second question is the impact of noise. Noise means incorrectly generating the information (``type'' or ``scope'') and it will mislead the subsequent generation of the ``subject''. To verify this conjecture, we divided the test set into two parts: one that correctly predicted the information before the ``subject'' and the other with incorrect predictions. The results on each part are shown in \Tab\ref{tab:codetfive-mcmdjs-different-output-subject-split}. As shown in the figure, all difference scores in ``Prefix Correct'' are higher than those in ``Prefix Wrong'', which indicates that the model gains more when it can correctly predict the information before ``subject''. Moreover, when the model generates ``scope'', it also pays some attention to ``type'' as shown in \Fig~\ref{fig:attention-weight-scope}. As the performance of ``subject'' is based on the information before it and the performance of ``scope'' can also depend on the ``type'', training with the combination of ``type'' and ``scope'' makes better performance on each component in the commit message. Training using these two kinds of information can improve the performance of predicting both of them as shown in \Tab~\ref{tab:codetfive-mcmdjs-different-output}, so it is better to train with both rather than one of them.
}

\section{Method}

In this section, we elaborate on our approach for generating commit messages with the knowledge.
We first propose a commit knowledge model learning from data with type and scope information.
Then, the details of the commit message generation are elaborated.
At last, we present a novel dynamic denoising training method to learn with noisy commit knowledge.

\subsection{Commit Knowledge Model}
To introduce specific knowledge into models, a recent trend is to adopt the retrieval-based paradigm to retrieve knowledge from a knowledge graph \citep{LvZH20}.
However, since the retrieval-based paradigm heavily relies on a large-scale knowledge graph or labeled examples, some works are proposed based on a modeling-based paradigm. 
In this paradigm, neural knowledge models are built to memorize specific knowledge into its parameters during training \citep{BosselutRSMCC19}.
These knowledge models are built upon a pre-trained Transformer (e.g., GPT~\citep{radford2018improving}, BERT \citep{DevlinCLT19}) and fine-tuned on labeled data \citep{ZhouGSPZJ21} or triples of knowledge graph \citep{BosselutRSMCC19}.

Motivated by the modeling-based paradigm, we propose a Transformer-based commit knowledge model trained on dataset $\hat \gD$ with type and scope to encapsulate commit knowledge. 
Each example in $\hat \gD$ composes a code change $\hat c$ and a commit message consisting of type $\hat t$, scope $\hat s$ and subject $\hat x$.
Formally, given a $\hat c$ and $\hat x$, we respectively pass them into an encoder-decoder based Transformer
to generate the type and scope, i.e.,
\begin{align}
\mH &= \transformerenc(\hat c ;\theta^{(enc)}), \label{equ:enc} \\
\hat \vp_i &= \transformerdec(\hat x, \mH ;\theta^{(dec)}) \in \gV, \label{equ:dec}
\end{align}
where $\mH$ denotes hidden states from the encoder; 
$\gV$ denotes token vocabulary and $\hat \vp_i$ is a probability distribution over $\gV$. 
$\transformerenc(\cdot; \theta)$ and $\transformerdec(\cdot; \theta)$ stand for $\theta$-parameterized pre-trained Transformer encoder and decoder, respectively.

During training, we leverage a cross-entropy loss to optimize the commit knowledge model $\{\theta^{(enc)}, \theta^{(dec)}\}$, towards type and scope generation, which is defined as,
\begin{align}
 \gL^{(km)} = - \dfrac{1}{|\hat N|}\sum\nolimits_{i=1}^{\hat N}\log \hat \vp_{i}(y_i), \text{~~~where}~y_i \in \{\hat t,\hat s\} 
 \label{equ:loss_km},
\end{align}
where $\hat \vp_{i}(y_i)$ denotes fetching the probability of the $i$-th gold token $y_i \in \{\hat t,\hat s\}$ from $\hat \vp_{i}$, and $\hat N$ is length of gold type and scope.

The encoder input of the knowledge model is formatted as $<S><code \ change></S>$ and the decoder output is formatted as $<S><subject></S><type, scope></S>$ where $<S>$ means the start of each component(code change, type, scope or subject) sequence token. $</S>$ means the end of whole sequence token. $<subject>$ is given in the decoder. In this way, the knowledge model can use both $<code \ change>$ and $<subject>$ to generate $<type>$ and $<scope>$.

\paragraph{Large-Scale Data Labeling.} For data in large-scale dataset $\bar \gD$ without type and scope, our commit knowledge model can derive their type $\widetilde t$ and scope $\widetilde s$ based on their code change $\bar c$ and subject $\bar x$ in commit message through Equ.\ref{equ:enc} and Equ.\ref{equ:dec}.

\subsection{Commit Message Generation}\label{sec:CMG}
The dataset $\gD$ can be split into two parts: $\bar \gD$ which does not have type or scope in each commit message, and $\hat\gD$ is the other.
As we label type and scope for large-scale $\bar \gD$ with a high proportion in the total dataset by a well-trained commit knowledge model, all examples in the full dataset, $\gD = \bar \gD \cup \hat\gD$, compose a code change $c$ and a commit message consisting of type $t$, scope $s$ and subject $x$.
For a given code change $c$, we pass it into an encoder-decoder based Transformer to generate a commit message consisting of type, scope, and subject, i.e.,
\begin{align}
\vp_i &= \transformer(c ;\theta^{(m)}) \in \gV, 
\end{align}
where $\gV$ denotes token vocabulary and $\vp_i$ is a probability distribution over $\gV$. $\transformer(\cdot; \theta^{(m)})$ stands for pre-trained encoder-decoder based Transformer. For the commit message generator training, the loss function can be denoted as,
\begin{align}
 \gL^{(cmg)} = - \dfrac{1}{|N|}\sum\nolimits_{i=1}^{N}\log \vp_{i}(y_i), \text{~~~where}~y_i \in \{t, s, x\}
 \label{equ:loss_cmg},
\end{align}
where $\vp_{i}(y_i)$ is the probability of the $i$-th gold token $y_i \in \{t, s, x\}$ in $\vp_{i}$, and $N$ is length of gold type, scope and subject.

The encoder input of the knowledge model is formatted as $<S><code \ change></S>$ and the decoder output is formatted as $<S><type, scope></S><subject></S>$ where $<S>$ and $</S>$ means the same as above. All of the components in the decoder is not given and need to be generated.

\subsection{Dynamic Denoising Training}
However, since types and scopes are generated by the commit knowledge model that is trained on a subset of the dataset $\hat\gD$, it is inevitable that noise (i.e., error prediction) lurks in the generated types and scopes.
Recently, many works \citep{HanYYNXHTS18,Han20Survey} have proven that an effective denoising method can improve the model performance.

In this work, each example $\hat e$ in the dataset $\hat \gD$ has labeled type and scope in its commit message, i.e.,
\begin{align}
\hat t &\sim P(\rt|\hat c, \hat x; \theta^{(human)}), \label{equ:human1}\\
\hat s &\sim P(\rs|\hat t, \hat c, \hat x; \theta^{(human)}), \label{equ:human2}
\end{align}
where $\theta^{(human)}$ denotes the human annotators.
For large-scale dataset $\bar \gD$, each example $\bar e$ has labeled type and scope by commit knowledge model, i.e.,
\begin{align}
\bar t &\sim P(\rt|\bar c, \bar x; \theta^{(enc)}, \theta^{(dec)}), \\
\bar s &\sim P(\rs|\bar t, \bar c, \bar x; \theta^{(enc)}, \theta^{(dec)}),
\end{align}
where $\theta^{(enc)}$ and $\theta^{(dec)}$ denote the encoder and decoder of the commit knowledge model. 

Since the commit knowledge model is learned from dataset $\hat \gD$, part of $\bar t$ and $\bar s$ are subject to distributions in Equ.\ref{equ:human1} and Equ.\ref{equ:human2}, respectively. To extrapolate to approximate distributions for clean and noisy data, we leverage the expectation–maximization (EM) algorithm \citep{dempster1977maximum} to deduce two distributions from training loss.

In addition, we propose a novel dynamic denoising training method that composes a dynamic distribution list and a distribution-aware confidence function.
During model training, we build and update the dynamic distribution list $\mL$ to record training loss calculated by Equ.\ref{equ:loss_cmg}.
At the beginning of each epoch, the distributions for clean and noisy data are re-deduced by the EM algorithm, i.e.,
\begin{align}
\mu, \nu = \emal(\mL)
\end{align}
where $\mu$ and $\nu$ denote approximate distributions for clean and noisy data, respectively.

To assign smaller weights to noisy data and greater weights to clean data, we reformalize training loss by the distribution-aware confidence function based on these two distributions, i.e.,
\begin{align}
    \gC(l_i) = \dfrac{P(l_i|\mu) + \alpha^{-l_i} \times P(l_i|\nu)}{P(l_i|\mu) + P(l_i|\nu)},\label{equ:alpha}
    \text{~~~where}~l_i \in \mL
\end{align}
\begin{align}
    \gL^{(dsc)} = - \gC(l_i) \dfrac{1}{|N|}\sum\nolimits_{i=1}^{N}\log \vp_{i}(y_i), 
    \text{~~~where}~y_i \in \{t, s, x\} \label{equ:loss_dsc}, \text{and}~l_i \in \mL
\end{align}
where $\gC$ and $\gL^{(dsc)}$ denote the distribution-aware confidence and reformalized training loss. $\alpha$ is a hyperparameter to compensate the bias of the EM algorithm.

\section{Experiments}

\subsection{Dataset}\label{sec:dataset}

\mcmd is selected because it is by far the largest peer-reviewed commit message dataset.
Previous studies~\citep{JiangAM17, LoyolaMM17, LiuXHLXW18, LiuLZFDQ19, NieGZLLX21} compared their methods with others under the dataset which is from Java repositories. 

To better validate our methods under different programming languages' repositories, we compare our method with others under all five programming languages' subsets of \mcmd including \mcmdjs, \mcmdcsharp, \mcmdpython, \mcmdcpp, and \mcmdjava. Each programming language (PL) subset of \mcmd (such as \mcmdjs) which is denoted as \mcmdpl has 450,000 pairs of code changes and commit messages. In each PL subset (i.e., \mcmdpl), 360,000 / 45,000 / 45,000 commits were randomly selected as training, valid, and test set.

Each \mcmdpl can be split into two parts: one part does not have type or scope in each commit message (as described $\bar \gD$ in \Sec\ref{sec:CMG}) and the other part does (as described $\hat \gD$). 
The former contains commit messages which are \textbf{u}nmatched with the AngularJS rule so it is denoted as \mcmdplu. The latter has rule-\textbf{m}atched commit messages so it is denoted as \mcmdplm. $\mcmdpl = \mcmdplm \cup \mcmdplu$.
For example, \mcmdjsm is the subset of \mcmdjs and it has 31,213 / 3,976 / 3,976 commits in the training, valid, and test set. $\mcmdjs = \mcmdjsm \cup \mcmdjsu$. Both \mcmdjsu and \mcmdjsm can be used to evaluate the performance of generated subject component while only \mcmdjsm can be used to evaluate the type and scope components.

\revise{Considering that different splitting strategies also influence performance, we conduct experiments under the setting of splitting by time to evaluate the robustness of our model. And we use the same setting of splitting by time as the paper~\citep{TaoWSDH0ZZ21}.}

\revise{Please note that both the baseline models and our proposed approach initiate training with the identical training set from MCMD. In our methodology, the knowledge model undergoes training on a subset of the original training set, generating type and scope for samples where (type, scope) information is absent in the initial training set. Subsequently, the knowledge model augments the training set, which is then utilized for our denoising training procedure.}

\subsection{Metrics}
Three widely-used metrics (BLEU, \meteor, and \rougel) are used to evaluate the similarity between the generation and the reference. 
BLEU~\citep{PapineniRWZ02} calculates the average of the modified n-gram precision to measure the precision. According to the human study~\citep{TaoWSDHZZZ22}, \bleunorm is the most consistent BLEU variant with human judgments on the quality of commit messages so it is selected. \meteor~\citep{BanerjeeL05} computes the harmonic mean of unigram precision and unigram recall of the generated results against the ground truth. \rougel~\citep{lin-2004-rouge} calculates the F-score of precision and recall based on the longest common sub-sequences between the generation and the ground truth.

As our model has the ability to generate type and scope, the F1 score is chosen to evaluate these two components because all of the ``type'' and more than 74\% of the ``scope'' have only one token, which is too short to use metrics designed for sentence. F1 score is interpreted as a harmonic mean of the precision and recall~\citep{sasaki2007truth}.

Moreover, as automatic metrics ``are not reliable enough to replace human evaluation for code documentation generation tasks''~\citep{HuCWXLZ22}, we also conduct a human evaluation. The details are described at \Sec~\ref{sec:human-eval}.

\subsection{Experimental Settings}
We choose five state-of-the-art commit message generation methods, i.e., \Commitgen~\citep{JiangAM17}, \Nmt~\citep{LoyolaMM17}, \Nngen~\citep{LiuXHLXW18}, \Ptrnet~\citep{LiuLZFDQ19}, and \Corec~\citep{WangXLHWG21} to compare with our model, \Ourmethod.  
\Atom~\citep{Liu20Atom} and \Fira~\citep{DongLZ0LZ022} take Abstract Syntax Trees from Java files to help the commit message generation and they cannot be directly migrated to the experimental dataset in which most of the files are multi-programming-language. Therefore, these two models are not selected in our comparison.
All of the baselines' reproduction follows their reproducible repository and the description in their paper. Moreover, we use the weights of CodeT5~\citep{0034WJH21} to initialize our model. For the optimizer, we use AdamW~\citep{LoshchilovH19} with the learning rate 5e-5. The batch size is 64, and the max number of epochs is 30. Most of the experiments are conducted on a server with 2 GPUs of NVIDIA Tesla V100 and it takes about 40 minutes each epoch for our model including training and validation.

\revise{Our method, leveraging the knowledge model, can generate additional information (namely, ``type'' and ``scope'') compared to other baselines. To ensure a fair comparison, we remove the ``type'' and ``scope'' parts from each commit message generated by our model to compare with the reference under \mcmdplu of the test set, due to the reference without ``type'' and ``scope''.} On the other hand, for \mcmdplm, a dataset with ``type'' and ``scope'' for each reference, all parts of each sentence in the generation results are compared with the reference.
Although previous baselines are evaluated on one programming language dataset in their papers, we use five programming languages' subsets of \mcmd to compare these models, which makes our conclusion more reliable.

\revise{In addition to random splitting, we also experimented with another splitting strategy: splitting by time. During the training of the knowledge model in that strategy, we find that older commits did less following the good practice.
For example, in the \mcmdjs under this splitting strategy, only 55.23\% of the samples in the training set with a quantity ratio of 80\% follow good practice.
This trend suggests an increasing adoption of best practices over time. 
Meanwhile, the samples having scope information in the training set under this setting are too small to train a knowledge model for other programming languages except for JavaScript. Specifically, there are only 128 samples having scope information in \mcmdjava. To deal with this issue, we use the knowledge model trained on \mcmdjsm (also under the setting of split-by-time) for other languages. 
This setup can also be used in real development situations.
Other details are the same as training on the dataset split randomly.
}

To make a deep analysis, we take \mcmdjs as an example to compare the ablation model and evaluate the generation performance of ``type'' and ``scope''. These experimental results show the ability of each component in our model and different-aspect abilities. Considering the limitation of automatic metrics, we also made a human evaluation to make our comparison results more in line with human standards. 
Moreover, we also selected some cases to illustrate the differences in the effects of different methods.

\begin{table*}[!t]
\centering
\caption{Model performance on the test set of \mcmd.}
\label{tab:baselines-comparison-all}
\begin{tabular}{llcccccc}
\toprule
\textbf{Dataset} & \bf Metric & \bf \Commitgen    & \bf \Nmt  & \bf \Nngen    & \bf \Ptrnet   & \bf \Corec    & \bf \Ourmethod    \\
\cmidrule(r){1-2} \cmidrule(r){3-8}
\multirow{3}{*}{\mcmdjs}
& \bf \bleu	& 17.40	& 17.08	& 18.03	& 19.59	& 19.84	& \bf{24.22} $\uparrow22.10\%$\\
& \bf \meteor	& 20.50	& 21.13	& 22.46	& 24.61	& 23.84	& \bf{28.55} $\uparrow16.00\%$\\
& \bf \rouge	& 19.94	& 20.54	& 21.27	& 24.60	& 23.36	& \bf{29.14} $\uparrow18.45\%$\\
\cmidrule(r){1-8}

\multirow{3}{*}{\mcmdcsharp}
& \bf \bleu	& 18.15	& 17.32	& 22.91	& 19.72	& 22.23	& \bf{24.72} $\uparrow7.88\%$\\
& \bf \meteor	& 20.18	& 19.81	& 26.22	& 22.33	& 25.38	& \bf{27.00} $\uparrow2.99\%$\\
& \bf \rouge	& 19.32	& 20.02	& 24.79	& 21.99	& 24.87	& \bf{27.77} $\uparrow11.68\%$\\
\cmidrule(r){1-8}

\multirow{3}{*}{\mcmdpython}
& \bf \bleu	& 11.10	& 11.52	& 16.64	& 15.99	& 15.13	& \bf{20.01} $\uparrow20.26\%$\\
& \bf \meteor	& 15.17	& 16.40	& 20.84	& 21.18	& 20.29	& \bf{24.07} $\uparrow13.62\%$\\
& \bf \rouge	& 13.01	& 14.41	& 19.44	& 20.76	& 18.81	& \bf{25.53} $\uparrow22.98\%$\\
\cmidrule(r){1-8}

\multirow{3}{*}{\mcmdcpp}
& \bf \bleu	& 11.58	& 11.56	& 13.69	& 13.07	& 13.80	& \bf{18.20} $\uparrow31.93\%$\\
& \bf \meteor	& 14.61	& 14.75	& 17.18	& 16.86	& 17.42	& \bf{20.93} $\uparrow20.12\%$\\
& \bf \rouge	& 13.53	& 14.04	& 16.25	& 17.08	& 16.62	& \bf{22.74} $\uparrow33.11\%$\\
\cmidrule(r){1-8}

\multirow{3}{*}{\mcmdjava}
& \bf \bleu	& 12.39	& 13.39	& 17.81	& 15.33	& 16.09	& \bf{19.81} $\uparrow11.21\%$\\
& \bf \meteor	& 14.16	& 16.00	& 22.12	& 19.13	& 19.58	& \bf{22.33} $\uparrow0.96\%$\\
& \bf \rouge	& 12.94	& 15.33	& 20.87	& 18.64	& 18.67	& \bf{23.31} $\uparrow11.71\%$\\
\toprule

\multirow{3}{*}{\bf Overall}
& \bf \bleu	& 14.12	& 14.17	& 17.82	& 16.74	& 17.42	& \bf{21.39} $\uparrow20.07\%$\\
& \bf \meteor	& 16.92	& 17.62	& 21.76	& 20.82	& 21.30	& \bf{24.58} $\uparrow12.92\%$\\
& \bf \rouge	& 15.75	& 16.87	& 20.52	& 20.61	& 20.47	& \bf{25.70} $\uparrow24.66\%$\\

\bottomrule
\end{tabular}
\end{table*}

\begin{table*}[!t]
\centering
\small
\caption{Model performance on each subset of \mcmd test set.}
\label{tab:baselines-comparison-all-split-all2sub}
\begin{tabular}{lllcccccc}
\toprule
\bf PL & \bf Dataset & \bf Metric & \bf \Commitgen    & \bf \Nmt  & \bf \Nngen    & \bf \Ptrnet   & \bf \Corec    & \bf \Ourmethod    \\
\cmidrule(r){1-3} \cmidrule(r){4-9}

\multirow{6}{*}{\rotatebox{90}{JavaScript}}
& \multirow{3}{*}{\shortstack{\mcmdjsu \\ (41024)}}
& \bf \bleu	& 16.29	& 16.07	& 17.12	& 18.62	& 18.86	& \bf{22.86} $\uparrow21.18\%$\\
&& \bf \meteor	& 18.91	& 19.64	& 20.95	& 23.13	& 22.40	& \bf{26.35} $\uparrow13.91\%$\\
&& \bf \rouge	& 18.92	& 19.59	& 20.44	& 23.59	& 22.47	& \bf{27.77} $\uparrow17.71\%$\\
\cmidrule(r){2-9}
& \multirow{3}{*}{\shortstack{\mcmdjsm \\ (3976)}}
& \bf \bleu	& 28.92	& 27.49	& 27.39	& 29.53	& 29.88	& \bf{38.28} $\uparrow28.11\%$\\
&& \bf \meteor	& 36.97	& 36.54	& 38.00	& 39.92	& 38.78	& \bf{51.28} $\uparrow28.48\%$\\
&& \bf \rouge	& 30.53	& 30.37	& 29.84	& 35.01	& 32.62	& \bf{43.26} $\uparrow23.58\%$\\
\cmidrule(r){1-9}

\multirow{6}{*}{\rotatebox{90}{C\#}}
& \multirow{3}{*}{\shortstack{\mcmdcsu \\ (44646)}}
& \bf \bleu	& 18.24	& 17.37	& 22.96	& 19.75	& 22.29	& \bf{24.72} $\uparrow7.67\%$\\
&& \bf \meteor	& 20.30	& 19.85	& 26.24	& 22.35	& 25.43	& \bf{26.93} $\uparrow2.62\%$\\
&& \bf \rouge	& 19.41	& 20.07	& 24.83	& 22.01	& 24.92	& \bf{27.75} $\uparrow11.34\%$\\
\cmidrule(r){2-9}
& \multirow{3}{*}{\shortstack{\mcmdcsm \\ (354)}}
& \bf \bleu	& 6.25	& 10.92	& 17.46	& 15.90	& 14.83	& \bf{24.95} $\uparrow42.89\%$\\
&& \bf \meteor	& 5.86	& 13.67	& 23.48	& 20.28	& 18.57	& \bf{36.30} $\uparrow54.60\%$\\
&& \bf \rouge	& 7.42	& 13.68	& 20.76	& 19.11	& 17.93	& \bf{30.67} $\uparrow47.76\%$\\
\cmidrule(r){1-9}

\multirow{6}{*}{\rotatebox{90}{Python}}
& \multirow{3}{*}{\shortstack{\mcmdpyu \\ (44646)}}
& \bf \bleu	& 11.10	& 11.52	& 16.64	& 16.00	& 15.13	& \bf{19.99} $\uparrow20.14\%$\\
&& \bf \meteor	& 15.13	& 16.37	& 20.79	& 21.14	& 20.25	& \bf{23.95} $\uparrow13.27\%$\\
&& \bf \rouge	& 13.01	& 14.42	& 19.43	& 20.74	& 18.81	& \bf{25.46} $\uparrow22.74\%$\\
\cmidrule(r){2-9}
& \multirow{3}{*}{\shortstack{\mcmdpym \\ (354)}}
& \bf \bleu	& 11.43	& 11.34	& 16.78	& 15.47	& 15.28	& \bf{22.61} $\uparrow34.74\%$\\
&& \bf \meteor	& 19.53	& 20.87	& 28.27	& 26.40	& 25.39	& \bf{39.13} $\uparrow38.43\%$\\
&& \bf \rouge	& 12.74	& 13.00	& 20.41	& 22.69	& 19.48	& \bf{34.05} $\uparrow50.02\%$\\
\cmidrule(r){1-9}

\multirow{6}{*}{\rotatebox{90}{C++}}
& \multirow{3}{*}{\shortstack{\mcmdcppu \\ (44681)}}
& \bf \bleu	& 11.61	& 11.59	& 13.69	& 13.09	& 13.83	& \bf{18.21} $\uparrow31.74\%$\\
&& \bf \meteor	& 14.64	& 14.78	& 17.15	& 16.87	& 17.43	& \bf{20.85} $\uparrow19.65\%$\\
&& \bf \rouge	& 13.57	& 14.09	& 16.25	& 17.09	& 16.65	& \bf{22.72} $\uparrow32.92\%$\\
\cmidrule(r){2-9}
& \multirow{3}{*}{\shortstack{\mcmdcppm \\ (319)}}
& \bf \bleu	& 6.72	& 6.73	& 13.89	& 10.22	& 9.99	& \bf{16.94} $\uparrow21.98\%$\\
&& \bf \meteor	& 10.57	& 10.37	& 21.87	& 15.51	& 16.35	& \bf{31.08} $\uparrow42.11\%$\\
&& \bf \rouge	& 7.99	& 7.04	& 16.53	& 15.79	& 11.87	& \bf{25.66} $\uparrow55.28\%$\\
\cmidrule(r){1-9}

\multirow{6}{*}{\rotatebox{90}{Java}}
& \multirow{3}{*}{\shortstack{\mcmdju \\ (44829)}}
& \bf \bleu	& 12.36	& 13.39	& 17.79	& 15.33	& 16.09	& \bf{19.79} $\uparrow11.21\%$\\
&& \bf \meteor	& 14.12	& 15.99	& 22.09	& 19.13	& 19.57	& \bf{22.28} $\uparrow0.89\%$\\
&& \bf \rouge	& 12.91	& 15.32	& 20.85	& 18.63	& 18.66	& \bf{23.28} $\uparrow11.67\%$\\
\cmidrule(r){2-9}
& \multirow{3}{*}{\shortstack{\mcmdjm \\ (171)}}
& \bf \bleu	& 19.77	& 14.01	& 23.03	& 15.17	& 16.47	& \bf{25.27} $\uparrow9.70\%$\\
&& \bf \meteor	& 24.29	& 18.40	& 29.85	& 18.91	& 22.02	& \bf{34.20} $\uparrow14.59\%$\\
&& \bf \rouge	& 21.72	& 18.31	& 26.70	& 20.13	& 20.14	& \bf{32.02} $\uparrow19.93\%$\\
\toprule

\multirow{6}{*}{\rotatebox{90}{\bf Overall}}
& \multirow{3}{*}{\shortstack{\mcmdu \\ (219826)}}
& \bf \bleu	& 13.88	& 13.95	& 17.65	& 16.52	& 17.21	& \bf{21.08} $\uparrow19.47\%$\\
&& \bf \meteor	& 16.58	& 17.29	& 21.45	& 20.48	& 20.99	& \bf{24.03} $\uparrow12.04\%$\\
&& \bf \rouge	& 15.51	& 16.65	& 20.36	& 20.36	& 20.26	& \bf{25.35} $\uparrow24.53\%$\\
\cmidrule(r){2-9}
& \multirow{3}{*}{\shortstack{\mcmdm \\ (5174)}}
& \bf \bleu	& 24.50	& 23.53	& 25.01	& 25.97	& 26.18	& \bf{34.55} $\uparrow31.96\%$\\
&& \bf \meteor	& 31.60	& 31.69	& 35.08	& 35.45	& 34.55	& \bf{47.62} $\uparrow34.33\%$\\
&& \bf \rouge	& 26.05	& 26.21	& 27.65	& 31.40	& 29.02	& \bf{40.31} $\uparrow28.39\%$\\

\bottomrule
\end{tabular}
\end{table*}

\begin{figure*}
    \includegraphics[width=0.48\columnwidth]{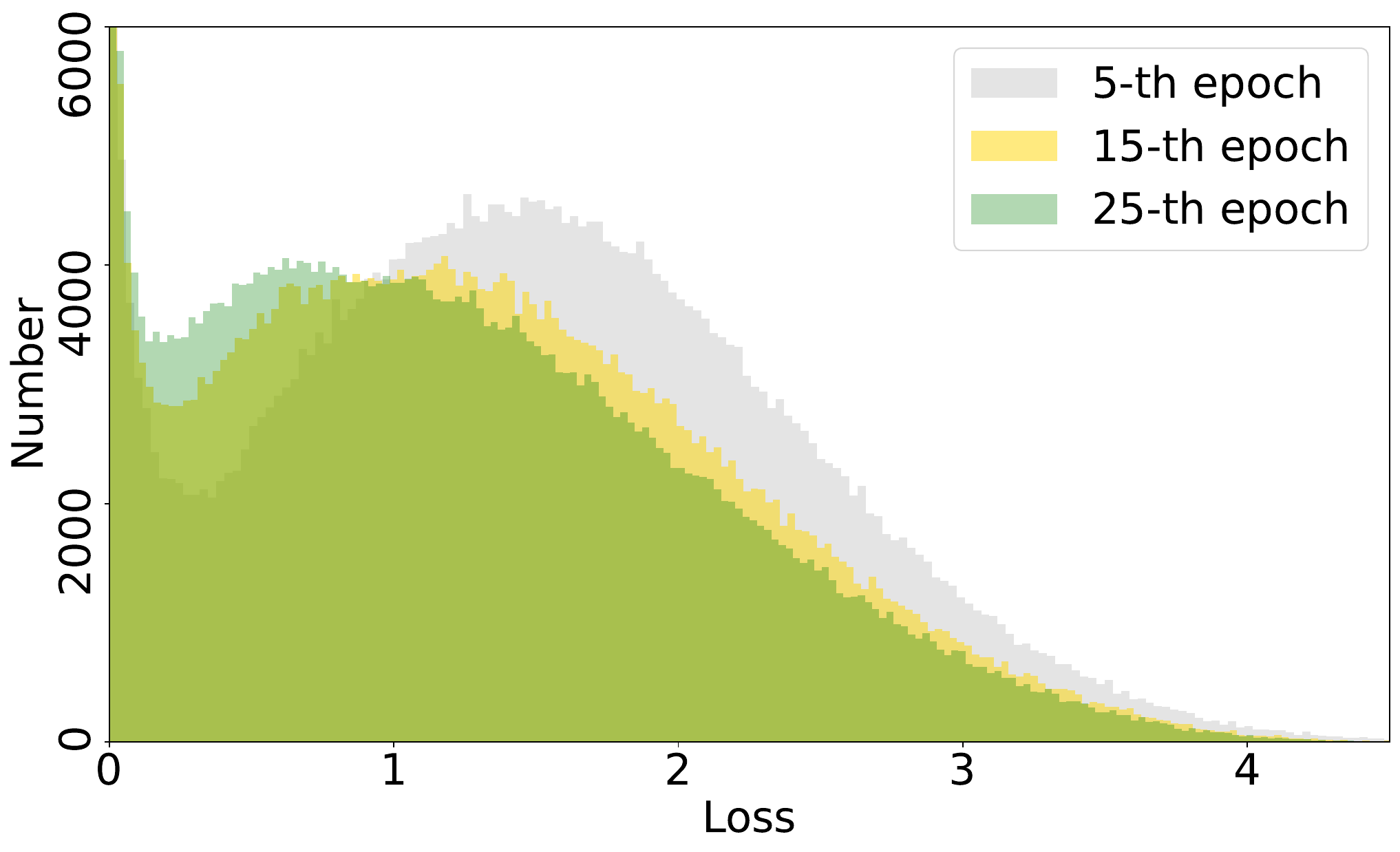}\label{b}
    \includegraphics[width=0.48\columnwidth]{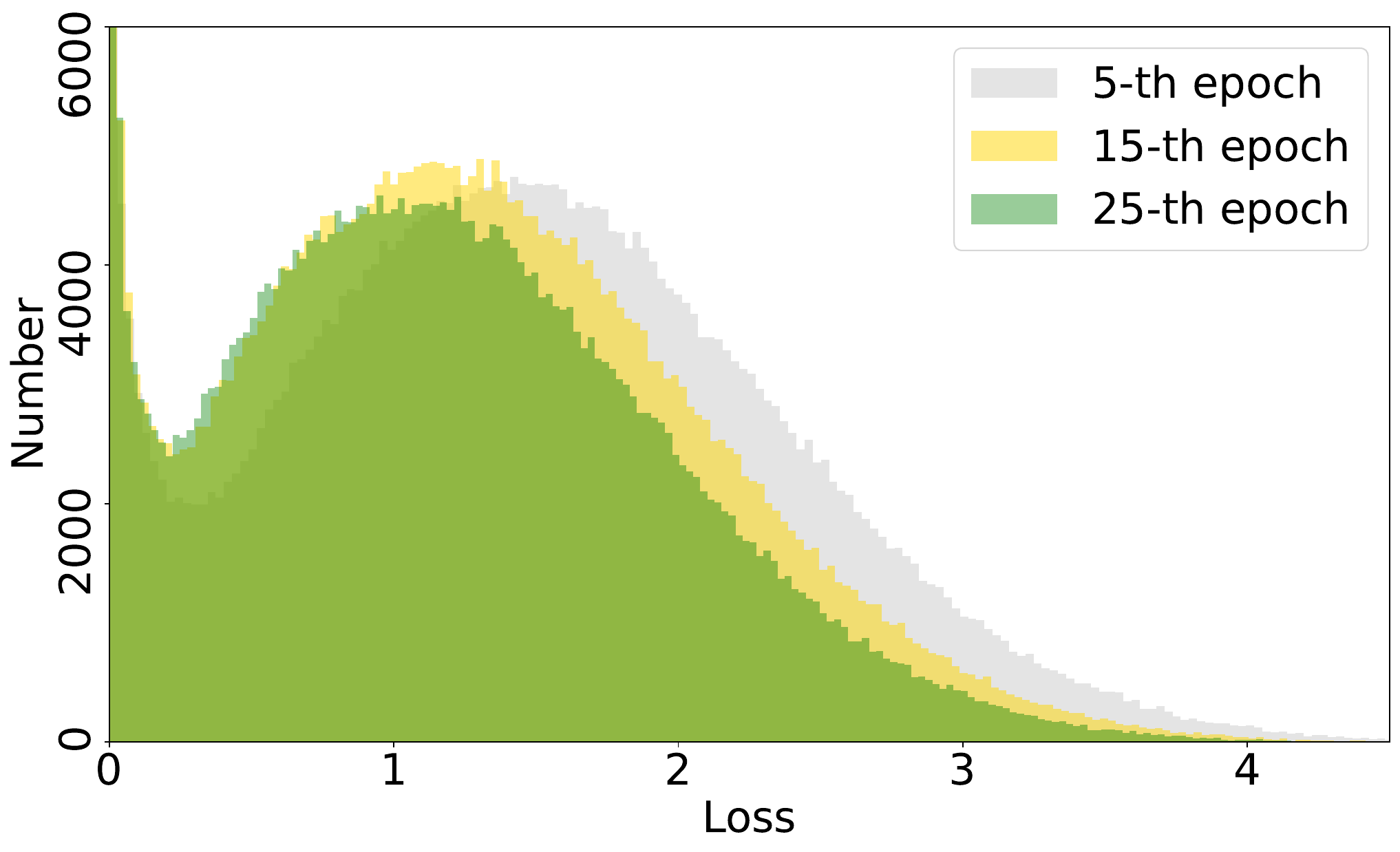}\label{a}
    \includegraphics[width=0.48\columnwidth]{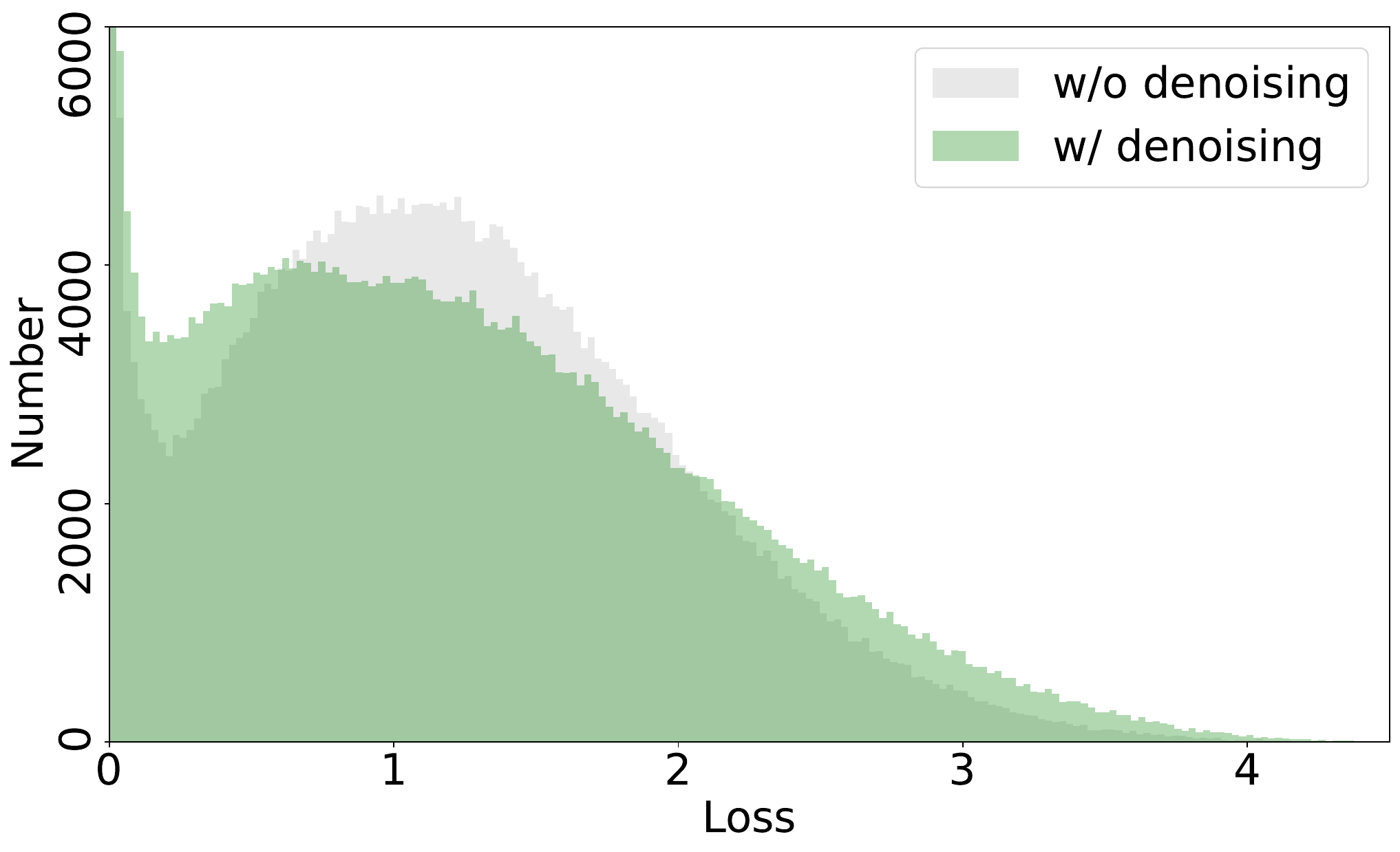}\label{c}
    \caption{Comparison of our method  (upper left) and it w/o denoising (upper right) on training loss distribution among epoch evolution; their comparison (down) on 25-th epoch.}
    \label{fig:denoising-change-distribution}
\end{figure*}

\subsection{Performance Comparison}\label{sec:perf}

\Tab~\ref{tab:baselines-comparison-all} shows the experimental results of all baselines and our model on \mcmd.
The overall compared results show that our method achieves better scores than baselines under all metrics (\bleunorm, \meteor, and \rougel) and leads the previous state-of-the-art baseline by 20.07\%, 12.92\%, and 24.66\% respectively. These scores validate the effectiveness and advancement of our method, KADEL, in generating the commit message.

\subsubsection{Performance on Each PL}

The magnitude of improvement varies across different PL subsets of \mcmd.

From the perspective of \bleunorm, the compared results on \mcmdjs, \mcmdpython and \mcmdcpp show that the improvement is more than 20\% and the results on \mcmdcsharp and \mcmdjava show about 10\%.

Although the \meteor score's improvement of our model on \mcmdjava is relatively small (less than 1\%), its \bleunorm and \rougel scores' improvement are significant. 
The reason why the \meteor score's improvement on \mcmdjava is small is probably that the commit knowledge model depends on the dataset with the type and scope information (i.e. \mcmdjm), and the proportion of \mcmdjm in \mcmdjava is such low as 0.40\%. To deal with this shortcoming, we also provide an improved solution (described in \Sec~\ref{sec:solution4java} )
\rougel scores of our model on other PLs' datasets show the improvement to the previous state-of-the-art one is from 11.68\% to 33.11\%.

Another finding is that generally the more \mcmdplm accounts for in \mcmdpl, the higher score of our model in each metric. This finding also validates the value of \mcmdplm and implies that our model takes full advantage of the commit knowledge in them.

\subsubsection{Performance on Each Subset}
As described in \Sec\ref{sec:dataset}, each PL's dataset can be split into two parts: \mcmdplu and \mcmdplm. AngualrJS commit rule is not required to follow for developers on the commits in \mcmdplu while it is required in \mcmdplm. \mcmdplu represents the commits in most of the repositories while \mcmdplu represents a few. The number of examples in the test set is written in parentheses for each item in the ``Dataset'' column.

To investigate the performance of our model in these different situations, we evaluate our method on each \mcmdplu and \mcmdplm \revise{of the test set}. 
The overall compared results on \mcmdu show that our method improves the \bleunorm, \meteor, \rougel scores by at least 19.47\%, 12.04\%, and 24.53\% than others respectively, which means that our method can be effectively applied to the most situation (about 97.70\% in \mcmd). For the situation which follows the AngularJS rule, our method shows far more advanced than others as results on \mcmdm. Compared with the previous state-of-the-art baseline, our method improves the \bleunorm, \meteor, \rougel scores by 31.96\%, 34.33\%, and 28.39\% respectively.

PL also influences the performance difference of the models on \mcmdplu and \mcmdplm.
For \mcmdjs, all models show consistently better performance on all metrics on \mcmdplm than on \mcmdplu.
For \mcmdcsharp, all models except ours show consistently better performance on all metrics on \mcmdplu than on \mcmdplm.
Similarly, for \mcmdpython and \mcmdjava, all models show better performance on all metrics on \mcmdplm than on \mcmdplu except \Nmt and \Ptrnet (only \bleu) on \mcmdpython, and \Ptrnet on \mcmdjava.
For \mcmdcpp, \Nngen and ours show better performance on most metrics on \mcmdplm than on \mcmdplu, and other models show the opposite. 
These findings indicate that there is a difference between the two subsets. In each \mcmdplu, our model shows the best performance among all metrics. 

\subsubsection{\revise{Performance on Splitting-By-Time}}
\revise{In this splitting strategy, the overall experimental results are shown in \Tab~\ref{tab:baselines-comparison-all-time}, and the performance of each subset is shown in \Tab~\ref{tab:baselines-comparison-all-split-all2sub-time}. As \Tab~\ref{tab:baselines-comparison-all-time} shows, our method outperforms baseline models across all metrics (\bleunorm, \meteor, and \rougel), surpassing the previous state-of-the-art by 25.37\%, 16.40\%, and 27.19\% respectively. These scores also validate the effectiveness and advancement of our method, KADEL. \Tab~\ref{tab:baselines-comparison-all-split-all2sub-time} also shows the scores have improved on each subset: from the perspective of \bleumosesnorm, the improvement ratio of our model ranges from 16.04\% to 26.93\% on \mcmdplu, which means that our method can be effectively applied to most situations (about 97.70\% in \mcmd). The substantial enhancements in performance reveal the capability of our knowledge model to facilitate knowledge transfer across time.
}

\begin{table*}[!t]
\centering
\caption{\revise{Model performance on the test set of \mcmd (Split by time).}}
\label{tab:baselines-comparison-all-time}
\begin{tabular}{llcccccc}
\toprule
\textbf{\revise{Dataset}} & \bf \revise{Metric} & \bf \revise{\Commitgen}    & \bf \revise{\Nmt}  & \bf \revise{\Nngen}    & \bf \revise{\Ptrnet}   & \bf \revise{\Corec}    & \bf \revise{\Ourmethod}    \\
\cmidrule(r){1-2} \cmidrule(r){3-8}

\multirow{3}{*}{ \revise{\mcmdjs} }
& \bf  \revise{\bleu} 	&  \revise{8.91} 	&  \revise{11.58} 	&  \revise{12.07} 	&  \revise{18.07} 	&  \revise{15.94} 	&  \revise{\bf{23.04} $\uparrow27.51\%$} \\
& \bf  \revise{\meteor} 	&  \revise{11.13} 	&  \revise{15.54} 	&  \revise{16.89} 	&  \revise{23.98} 	&  \revise{20.75} 	&  \revise{\bf{29.56} $\uparrow23.27\%$} \\
& \bf  \revise{\rouge} 	&  \revise{12.05} 	&  \revise{14.44} 	&  \revise{13.07} 	&  \revise{23.10} 	&  \revise{18.57} 	&  \revise{\bf{27.73} $\uparrow20.05\%$} \\
\cmidrule(r){1-8}

\multirow{3}{*}{ \revise{\mcmdcsharp} }
& \bf  \revise{\bleu} 	&  \revise{4.53} 	&  \revise{5.15} 	&  \revise{7.83} 	&  \revise{9.38} 	&  \revise{9.16} 	&  \revise{\bf{11.32} $\uparrow20.70\%$} \\
& \bf  \revise{\meteor} 	&  \revise{5.97} 	&  \revise{7.71} 	&  \revise{10.45} 	&  \revise{12.18} 	&  \revise{11.58} 	&  \revise{\bf{17.42} $\uparrow43.06\%$} \\
& \bf  \revise{\rouge} 	&  \revise{7.04} 	&  \revise{8.42} 	&  \revise{8.67} 	&  \revise{11.55} 	&  \revise{11.32} 	&  \revise{\bf{14.79} $\uparrow28.10\%$} \\
\cmidrule(r){1-8}

\multirow{3}{*}{ \revise{\mcmdpython} }
& \bf  \revise{\bleu} 	&  \revise{5.50} 	&  \revise{7.31} 	&  \revise{9.36} 	&  \revise{13.21} 	&  \revise{11.07} 	&  \revise{\bf{16.77} $\uparrow26.94\%$} \\
& \bf  \revise{\meteor} 	&  \revise{6.71} 	&  \revise{11.17} 	&  \revise{13.87} 	&  \revise{20.01} 	&  \revise{16.28} 	&  \revise{\bf{20.65} $\uparrow3.22\%$} \\
& \bf  \revise{\rouge} 	&  \revise{7.60} 	&  \revise{8.42} 	&  \revise{9.73} 	&  \revise{17.02} 	&  \revise{12.84} 	&  \revise{\bf{22.57} $\uparrow32.61\%$} \\
\cmidrule(r){1-8}

\multirow{3}{*}{ \revise{\mcmdcpp} }
& \bf  \revise{\bleu} 	&  \revise{7.08} 	&  \revise{8.52} 	&  \revise{9.30} 	&  \revise{10.94} 	&  \revise{11.72} 	&  \revise{\bf{14.55} $\uparrow24.10\%$} \\
& \bf  \revise{\meteor} 	&  \revise{8.96} 	&  \revise{11.10} 	&  \revise{12.29} 	&  \revise{14.33} 	&  \revise{15.40} 	&  \revise{\bf{16.55} $\uparrow7.47\%$} \\
& \bf  \revise{\rouge} 	&  \revise{9.80} 	&  \revise{10.60} 	&  \revise{10.53} 	&  \revise{13.50} 	&  \revise{14.33} 	&  \revise{\bf{19.06} $\uparrow33.04\%$} \\
\cmidrule(r){1-8}

\multirow{3}{*}{ \revise{\mcmdjava} }
& \bf  \revise{\bleu} 	&  \revise{8.08} 	&  \revise{9.49} 	&  \revise{10.73} 	&  \revise{13.30} 	&  \revise{12.93} 	&  \revise{\bf{15.68} $\uparrow17.87\%$} \\
& \bf  \revise{\meteor} 	&  \revise{9.23} 	&  \revise{13.86} 	&  \revise{14.34} 	&  \revise{17.55} 	&  \revise{16.61} 	&  \revise{\bf{18.30} $\uparrow4.28\%$} \\
& \bf  \revise{\rouge} 	&  \revise{8.74} 	&  \revise{11.13} 	&  \revise{11.57} 	&  \revise{15.71} 	&  \revise{14.37} 	&  \revise{\bf{18.72} $\uparrow19.14\%$} \\
\toprule

\multirow{3}{*}{\bf  \revise{Overall}} 
& \bf  \revise{\bleu} 	&  \revise{6.82} 	&  \revise{8.41} 	&  \revise{9.86} 	&  \revise{12.98} 	&  \revise{12.17} 	&  \revise{\bf{16.27} $\uparrow25.37\%$} \\
& \bf  \revise{\meteor} 	&  \revise{8.40} 	&  \revise{11.88} 	&  \revise{13.57} 	&  \revise{17.61} 	&  \revise{16.12} 	&  \revise{\bf{20.50} $\uparrow16.40\%$} \\
& \bf  \revise{\rouge} 	&  \revise{9.04} 	&  \revise{10.60} 	&  \revise{10.72} 	&  \revise{16.18} 	&  \revise{14.29} 	&  \revise{\bf{20.57} $\uparrow27.19\%$} \\

\bottomrule
\end{tabular}
\end{table*}

\begin{table*}[!t]
\centering
\small
\caption{\revise{Model performance on each subset of \mcmd test set (Split by time).}}
\label{tab:baselines-comparison-all-split-all2sub-time}
\begin{tabular}{lllcccccc}
\toprule
\bf \revise{PL} & \bf \revise{Dataset} & \bf \revise{Metric} & \bf \revise{\Commitgen}    & \bf \revise{\Nmt}  & \bf \revise{\Nngen}    & \bf \revise{\Ptrnet}   & \bf \revise{\Corec}    & \bf \revise{\Ourmethod}    \\
\cmidrule(r){1-3} \cmidrule(r){4-9}

\multirow{6}{*}{\rotatebox{90}{ \revise{JavaScript} }}
& \multirow{3}{*}{\shortstack{ \revise{\mcmdjsu}  \\  \revise{(35864)} }}
& \bf  \revise{\bleu} 	&  \revise{8.72} 	&  \revise{12.02} 	&  \revise{11.43} 	&  \revise{17.36} 	&  \revise{15.33} 	&  \revise{\bf{21.17} $\uparrow21.98\%$} \\
&& \bf  \revise{\meteor} 	&  \revise{10.98} 	&  \revise{16.08} 	&  \revise{16.01} 	&  \revise{22.89} 	&  \revise{19.96} 	&  \revise{\bf{25.91} $\uparrow13.20\%$} \\
&& \bf  \revise{\rouge} 	&  \revise{11.57} 	&  \revise{15.03} 	&  \revise{12.31} 	&  \revise{21.68} 	&  \revise{17.58} 	&  \revise{\bf{25.45} $\uparrow17.43\%$} \\
\cmidrule(r){2-9}
& \multirow{3}{*}{\shortstack{ \revise{\mcmdjsm}  \\  \revise{(9136)} }}
& \bf  \revise{\bleu} 	&  \revise{9.69} 	&  \revise{9.86} 	&  \revise{14.57} 	&  \revise{20.87} 	&  \revise{18.30} 	&  \revise{\bf{30.39} $\uparrow45.59\%$} \\
&& \bf  \revise{\meteor} 	&  \revise{11.72} 	&  \revise{13.41} 	&  \revise{20.33} 	&  \revise{28.25} 	&  \revise{23.87} 	&  \revise{\bf{43.87} $\uparrow55.30\%$} \\
&& \bf  \revise{\rouge} 	&  \revise{13.94} 	&  \revise{12.12} 	&  \revise{16.06} 	&  \revise{28.68} 	&  \revise{22.46} 	&  \revise{\bf{36.66} $\uparrow27.83\%$} \\
\cmidrule(r){1-9}

\multirow{6}{*}{\rotatebox{90}{ \revise{C\#} }}
& \multirow{3}{*}{\shortstack{ \revise{\mcmdcsu}  \\  \revise{(42352)} }}
& \bf  \revise{\bleu} 	&  \revise{4.73} 	&  \revise{5.36} 	&  \revise{8.08} 	&  \revise{9.57} 	&  \revise{9.52} 	&  \revise{\bf{11.11} $\uparrow16.04\%$} \\
&& \bf  \revise{\meteor} 	&  \revise{6.24} 	&  \revise{8.11} 	&  \revise{10.84} 	&  \revise{12.52} 	&  \revise{12.11} 	&  \revise{\bf{17.13} $\uparrow36.76\%$} \\
&& \bf  \revise{\rouge} 	&  \revise{7.37} 	&  \revise{8.77} 	&  \revise{8.95} 	&  \revise{11.70} 	&  \revise{11.70} 	&  \revise{\bf{14.95} $\uparrow27.69\%$} \\
\cmidrule(r){2-9}
& \multirow{3}{*}{\shortstack{ \revise{\mcmdcsm}  \\  \revise{(2648)} }}
& \bf  \revise{\bleu} 	&  \revise{1.26} 	&  \revise{1.76} 	&  \revise{3.86} 	&  \revise{6.23} 	&  \revise{3.36} 	&  \revise{\bf{14.66} $\uparrow135.23\%$} \\
&& \bf  \revise{\meteor} 	&  \revise{1.52} 	&  \revise{1.37} 	&  \revise{4.09} 	&  \revise{6.62} 	&  \revise{3.12} 	&  \revise{\bf{22.11} $\uparrow233.81\%$} \\
&& \bf  \revise{\rouge} 	&  \revise{1.79} 	&  \revise{2.74} 	&  \revise{4.22} 	&  \revise{9.11} 	&  \revise{5.21} 	&  \revise{\bf{12.31} $\uparrow35.11\%$} \\
\cmidrule(r){1-9}

\multirow{6}{*}{\rotatebox{90}{ \revise{Python} }}
& \multirow{3}{*}{\shortstack{ \revise{\mcmdpyu}  \\  \revise{(43677)} }}
& \bf  \revise{\bleu} 	&  \revise{5.61} 	&  \revise{7.34} 	&  \revise{9.32} 	&  \revise{13.21} 	&  \revise{11.06} 	&  \revise{\bf{16.77} $\uparrow26.93\%$} \\
&& \bf  \revise{\meteor} 	&  \revise{6.84} 	&  \revise{11.22} 	&  \revise{13.78} 	&  \revise{19.97} 	&  \revise{16.21} 	&  \revise{\bf{20.26} $\uparrow1.48\%$} \\
&& \bf  \revise{\rouge} 	&  \revise{7.73} 	&  \revise{8.54} 	&  \revise{9.79} 	&  \revise{17.08} 	&  \revise{12.93} 	&  \revise{\bf{22.59} $\uparrow32.22\%$} \\
\cmidrule(r){2-9}
& \multirow{3}{*}{\shortstack{ \revise{\mcmdpym}  \\  \revise{(1323)} }}
& \bf  \revise{\bleu} 	&  \revise{1.66} 	&  \revise{6.38} 	&  \revise{10.37} 	&  \revise{13.29} 	&  \revise{11.63} 	&  \revise{\bf{16.92} $\uparrow27.30\%$} \\
&& \bf  \revise{\meteor} 	&  \revise{2.30} 	&  \revise{9.71} 	&  \revise{16.62} 	&  \revise{21.32} 	&  \revise{18.49} 	&  \revise{\bf{33.51} $\uparrow57.15\%$} \\
&& \bf  \revise{\rouge} 	&  \revise{3.04} 	&  \revise{4.48} 	&  \revise{7.71} 	&  \revise{15.03} 	&  \revise{10.03} 	&  \revise{\bf{22.09} $\uparrow47.05\%$} \\
\cmidrule(r){1-9}

\multirow{6}{*}{\rotatebox{90}{ \revise{C++} }}
& \multirow{3}{*}{\shortstack{ \revise{\mcmdcppu}  \\  \revise{(44296)} }}
& \bf  \revise{\bleu} 	&  \revise{7.17} 	&  \revise{8.54} 	&  \revise{9.32} 	&  \revise{10.95} 	&  \revise{11.77} 	&  \revise{\bf{14.56} $\uparrow23.63\%$} \\
&& \bf  \revise{\meteor} 	&  \revise{9.08} 	&  \revise{11.13} 	&  \revise{12.28} 	&  \revise{14.35} 	&  \revise{15.41} 	&  \revise{\bf{16.43} $\uparrow6.66\%$} \\
&& \bf  \revise{\rouge} 	&  \revise{9.92} 	&  \revise{10.63} 	&  \revise{10.57} 	&  \revise{13.51} 	&  \revise{14.41} 	&  \revise{\bf{19.02} $\uparrow32.03\%$} \\
\cmidrule(r){2-9}
& \multirow{3}{*}{\shortstack{ \revise{\mcmdcppm}  \\  \revise{(704)} }}
& \bf  \revise{\bleu} 	&  \revise{1.56} 	&  \revise{7.67} 	&  \revise{8.28} 	&  \revise{9.91} 	&  \revise{8.63} 	&  \revise{\bf{14.20} $\uparrow43.29\%$} \\
&& \bf  \revise{\meteor} 	&  \revise{1.62} 	&  \revise{9.48} 	&  \revise{13.19} 	&  \revise{13.39} 	&  \revise{15.02} 	&  \revise{\bf{24.05} $\uparrow60.13\%$} \\
&& \bf  \revise{\rouge} 	&  \revise{2.60} 	&  \revise{8.66} 	&  \revise{8.46} 	&  \revise{12.99} 	&  \revise{9.36} 	&  \revise{\bf{21.59} $\uparrow66.14\%$} \\
\cmidrule(r){1-9}

\multirow{6}{*}{\rotatebox{90}{ \revise{Java} }}
& \multirow{3}{*}{\shortstack{ \revise{\mcmdju}  \\  \revise{(44456)} }}
& \bf  \revise{\bleu} 	&  \revise{8.14} 	&  \revise{9.55} 	&  \revise{10.77} 	&  \revise{13.35} 	&  \revise{13.02} 	&  \revise{\bf{15.64} $\uparrow17.14\%$} \\
&& \bf  \revise{\meteor} 	&  \revise{9.30} 	&  \revise{13.94} 	&  \revise{14.41} 	&  \revise{17.62} 	&  \revise{16.74} 	&  \revise{\bf{18.18} $\uparrow3.19\%$} \\
&& \bf  \revise{\rouge} 	&  \revise{8.79} 	&  \revise{11.16} 	&  \revise{11.61} 	&  \revise{15.73} 	&  \revise{14.46} 	&  \revise{\bf{18.65} $\uparrow18.59\%$} \\
\cmidrule(r){2-9}
& \multirow{3}{*}{\shortstack{ \revise{\mcmdjm}  \\  \revise{(544)} }}
& \bf  \revise{\bleu} 	&  \revise{3.15} 	&  \revise{5.29} 	&  \revise{7.34} 	&  \revise{9.65} 	&  \revise{5.71} 	&  \revise{\bf{19.40} $\uparrow101.00\%$} \\
&& \bf  \revise{\meteor} 	&  \revise{3.47} 	&  \revise{6.72} 	&  \revise{9.15} 	&  \revise{11.40} 	&  \revise{5.78} 	&  \revise{\bf{27.61} $\uparrow142.24\%$} \\
&& \bf  \revise{\rouge} 	&  \revise{4.60} 	&  \revise{8.81} 	&  \revise{8.88} 	&  \revise{14.31} 	&  \revise{7.24} 	&  \revise{\bf{24.14} $\uparrow68.71\%$} \\
\toprule

\multirow{6}{*}{\rotatebox{90}{\bf  \revise{Overall}}} 
& \multirow{3}{*}{\shortstack{ \revise{\mcmdu}  \\  \revise{(210645)} }}
& \bf  \revise{\bleu} 	&  \revise{6.83} 	&  \revise{8.45} 	&  \revise{9.74} 	&  \revise{12.74} 	&  \revise{12.04} 	&  \revise{\bf{15.68} $\uparrow23.05\%$} \\
&& \bf  \revise{\meteor} 	&  \revise{8.41} 	&  \revise{11.98} 	&  \revise{13.39} 	&  \revise{17.29} 	&  \revise{15.97} 	&  \revise{\bf{19.35} $\uparrow11.91\%$} \\
&& \bf  \revise{\rouge} 	&  \revise{8.99} 	&  \revise{10.68} 	&  \revise{10.60} 	&  \revise{15.75} 	&  \revise{14.11} 	&  \revise{\bf{19.96} $\uparrow26.76\%$} \\
\cmidrule(r){2-9}
& \multirow{3}{*}{\shortstack{ \revise{\mcmdm}  \\  \revise{(14355)} }}
& \bf  \revise{\bleu} 	&  \revise{6.75} 	&  \revise{7.76} 	&  \revise{11.62} 	&  \revise{16.51} 	&  \revise{13.98} 	&  \revise{\bf{25.03} $\uparrow51.64\%$} \\
&& \bf  \revise{\meteor} 	&  \revise{8.16} 	&  \revise{10.40} 	&  \revise{16.22} 	&  \revise{22.26} 	&  \revise{18.42} 	&  \revise{\bf{37.32} $\uparrow67.67\%$} \\
&& \bf  \revise{\rouge} 	&  \revise{9.78} 	&  \revise{9.39} 	&  \revise{12.46} 	&  \revise{22.50} 	&  \revise{16.92} 	&  \revise{\bf{29.61} $\uparrow31.63\%$} \\

\bottomrule
\end{tabular}
\end{table*}

\subsection{Ablation Study}

As our model has two components: one is to introduce commit knowledge, and another is for denoising training. We make an ablation study to investigate the value of each one. Training without knowledge means that the model is trained with the pairs of code changes and subjects in \mcmdpl. Training without the dynamic denoising module means that the model is trained with pairs of code changes and (type, scope, subject) discriminatively although there are many noisy pseudo labels (i.e., type and scope).

\Tab~\ref{tab:ablation-study} shows the commit message generated by the model trained without commit knowledge or without denoising module decreases the performance of our model. It indicates that each module is valuable to generate better commit messages.

\begin{table}[!t]
\centering
\caption{Abation performance in dataset \mcmdjs. BLEU is short for BLEU-Norm.}
\label{tab:ablation-study}
\begin{tabular}{lccc}
\toprule
\textbf{Model}  & \textbf{BLEU} & \textbf{\meteor}  & \textbf{\rougel}   \\  
\midrule
w/o Knowledge   & 20.72 	& 24.99 	& 26.17 \\
w/o Denoising   & 22.73 	& 27.28 	& 27.44 \\
\midrule
\multirow{1}{*}{Ours} & \textbf{24.22}  & \textbf{28.55}  & \textbf{29.14} \\ 
\bottomrule
\end{tabular}
\end{table}

\revise{
The incorporation of the pre-trained model contributes to the enhancement of our performance. When compared to other baselines, the pre-trained model, fine-tuned without additional knowledge, outperforms with a \bleunorm score of 20.72, surpassing the highest-scoring baseline by 0.88 points (19.84). Notably, our model achieves a \bleunorm score of 24.22, signifying a substantial advancement of 16.89\% over the 20.72 baseline, equating to an increase of 3.50. Therefore, our method's contribution, independent of the pre-trained model, accounts for approximately 80\% of the overall improvement.
}

\subsection{Analysis}

\subsubsection{Impact of Commit Knowledge Model.}

The performance of the commit knowledge model is shown in \Tab~\ref{tab:codetfive-enhanced-type_scope} in which the F1 score ranges from 60.46 to 76.17. 
The performance of the knowledge model training with <type> or <scope> is better than training with both two components. One way to decrease the noise is by choosing the knowledge model aimed at each component.
Although the commit knowledge model reaches a significant performance, using it to derive commit knowledge for large-scale data would still introduce noise because these F1 scores do not reach 100.

\begin{table}[!t]
\centering
\caption{The performance of our model to predict \type and \scope on \mcmdjsm in different settings of the decoder output. 
A checkmark means the data item is used for training in that setting (one row represents one setting).
}
\label{tab:codetfive-enhanced-type_scope}
\begin{tabular}{cccccccc}
\toprule
\multicolumn{4}{c}{\textbf{Training Setting}}  & \multicolumn{4}{c}{\textbf{Test Performance}} \\
\cmidrule(r){1-4} \cmidrule(r){5-8}
Input & \multicolumn{3}{c}{Decoder Output}    & \multicolumn{2}{c}{\revise{Type}} & \multicolumn{2}{c}{\revise{Scope}}  \\
\cmidrule(r){1-1} \cmidrule(r){2-4} \cmidrule(r){5-6} \cmidrule(r){7-8}
\cdiff & \revise{Subject} & \revise{Type} & \revise{Scope}  & \revise{EM}   & F1    & \revise{EM}   & F1   \\
\cmidrule(r){1-1} \cmidrule(r){2-4} \cmidrule(r){5-6} \cmidrule(r){7-8}
$\checkmark$ & $\checkmark$ & $\checkmark$  & $\checkmark$  & \revise{70.45}            & 70.12          & \revise{63.34}           & 60.46         \\ 
$\checkmark$ & $\checkmark$ & $\checkmark$  &               & \revise{\textbf{76.58}}   & \textbf{76.17} & \revise{-}               & -             \\
$\checkmark$ & $\checkmark$ &               & $\checkmark$  & \revise{-}                & -              & \revise{\textbf{65.70}}  & \textbf{63.50}\\
\toprule
\end{tabular}
\end{table}

\subsubsection{Impact of Denoising Training.}

\begin{figure}[!t]
\centering
    \includegraphics[width=0.8\columnwidth]{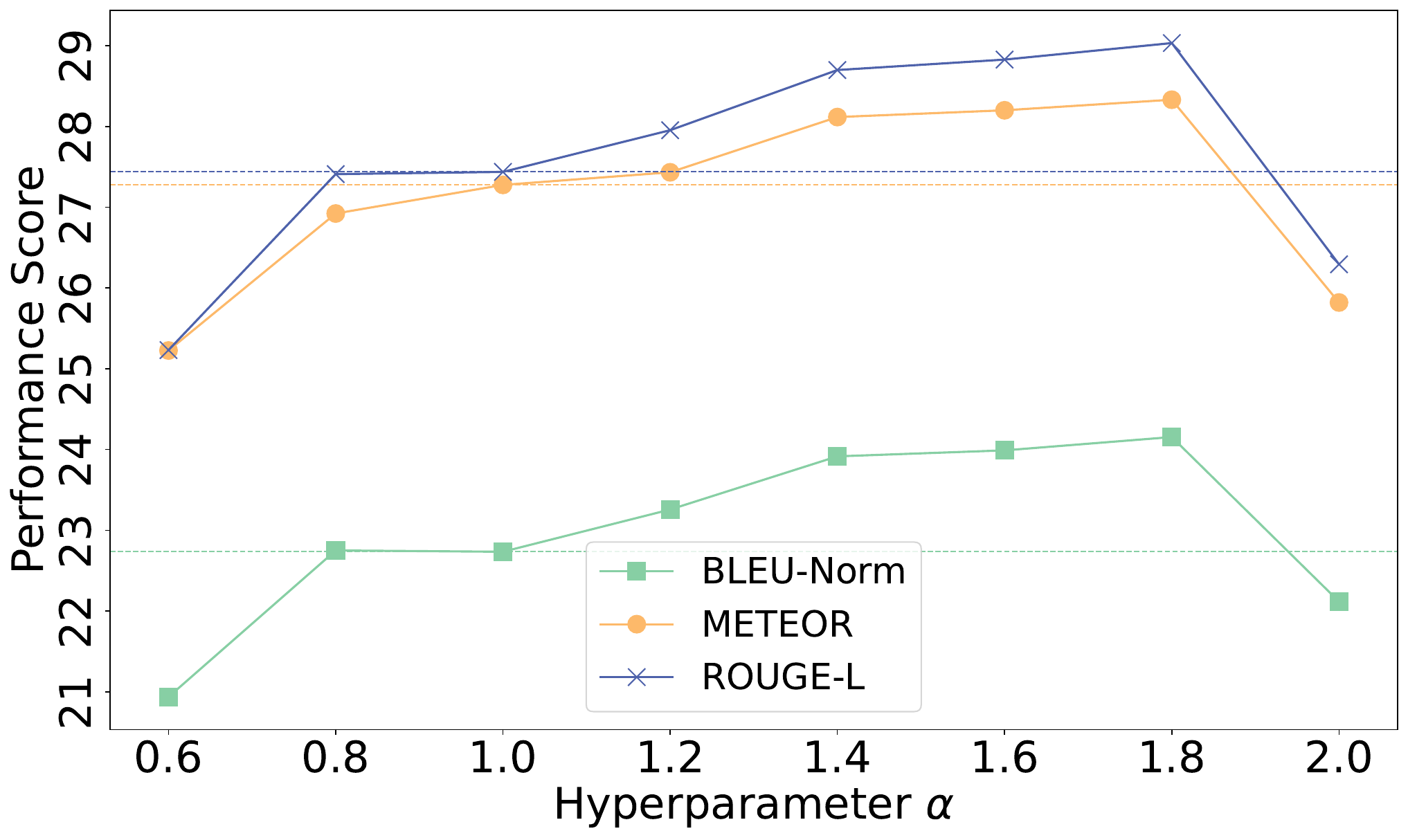}
    \caption{The performance under different hyperparameters $\alpha$ in different metrics. The dashed line represents the performance without denoising training.}
    \label{fig:hyperparameter}
\end{figure}

$\alpha$ is the hyperparameter to compensate for the bias of the EM algorithm as shown in Equ.~\ref{equ:alpha}. $\alpha$ equals one means that denoising is not used in the training. We select different values of hyperparameter and the corresponding performance results are shown in \Fig~\ref{fig:hyperparameter}. 
We find $1.8$ as the value of $\alpha$ according to its best score among others for \mcmdjs, and the great value leads to performance drops.
In addition, $\alpha$ less than $1.0$ means giving higher weight to data with generated type and scope, and the performance drops denote more noisy samples in noisy distribution predicted by EM.
The results demonstrate the effectiveness of our denoising training.
In addition, we investigate our denoising training by loss distribution evolution over epochs.
The ablation model without denoising is shown in the upper right subfigure of \Fig~\ref{fig:denoising-change-distribution}. Our method with denoising training can distinguish clean samples from noisy samples faster, as shown in the upper right subfigure of \Fig~\ref{fig:denoising-change-distribution}.
Comparing the upper left subfigure and lower subfigure of \Fig~\ref{fig:denoising-change-distribution}, it can find that denoising training can achieve more effective learning to push the loss distribution to move left faster.

\subsubsection{Human Evaluation.}\label{sec:human-eval}

\begin{table}[ht]
\footnotesize
\centering
\caption{The meaning of scores in human evaluation.}
\label{tab:human_score_meaning_3}
\begin{tabular}{c|l} 
\toprule
\multicolumn{2}{l}{\textbf{\underline{Content Adequacy}}}  \\
\multicolumn{2}{l}{Is the important information about the code changes reflected in the commit message?}  \\
\midrule
{0}   &  Missing all information about the code change. \\
\multirow{1}{*}{1}  &  Missing some important information that can hinder the understanding of the code changes.\\ 
\multirow{1}{*}{2}  &  Missing some information but some of the missing is not necessary to understand the code changes.   \\
\multirow{1}{*}{3}  &  Missing some info. but all missing is not necessary to understand the code changes.    \\
4   &   Not missing any information. \\
\midrule 
\multicolumn{2}{l}{\textbf{\underline{Conciseness}}}   \\
\multicolumn{2}{l}{Is there extraneous info. included in the commit message?}  \\
\midrule
{0}   &   {All of the information is unnecessary.    }\\
{1}   &   {Has a lot of unnecessary information. }\\
{2}   &   {Has some unnecessary information. }\\
{3}   &   {Has a little unnecessary information. }\\
{4}   &   {Has no unnecessary information.   }\\
\midrule 
\multicolumn{2}{l}{{\textbf{\underline{Expressiveness}}}}   \\
\multicolumn{2}{l}{{How readable and understandable is the commit message?}}  \\
\midrule 
{0}   &   {Cannot read and understand.} \\
{1}   &   {Is hard to read and understand.} \\
{2}   &   {Is somewhat readable and understandable.   } \\
{3}   &   {Is mostly readable and understandable. } \\
{4}   &   {Is easy to read and understand.} \\
\bottomrule
\end{tabular}
\end{table}

We conduct a human evaluation to compare the previous best baselines and our model. We randomly select $50$ commits from the test set of \mcmdjs and collect the corresponding commit message generated by previous best baselines (\Ptrnet and \Corec) and our model. Following best practices for human evaluation~\citep{LeeGMWK19}, three experts are invited to label the data manually. All of them have more than five years of programming experience. 
We define criteria in three aspects for manual labeling as shown in \Tab~\ref{tab:human_score_meaning_3} which shows the meaning of scores and is used to guide the raters to score following previous works~\citep{MorenoASMPV13, PanichellaPBZG16}.
According to the criteria, three raters give a score between 0 to 4 to measure the quality of each generated commit message in three aspects.
To evaluate the value of ``type'' and ``scope'' to the original commit message, all three raters also label the subject part in our model's generation results. To verify the agreement among the raters, we calculate the Kendall rank correlation coefficient values~\citep{kendall1945treatment}. The values of pairwise Kendall's Tau range from 0.71 to 0.93, which indicates that there is a high degree of agreement between the three raters and that scores are reliable.

\begin{table}[!t]
\centering 
\caption{Results of human evaluation (standard deviation in parentheses).}
\label{tab:human-evaluation}
\begin{tabular}{cccc}
\toprule
\textbf{Model}       & \textbf{Content Adequacy}        &\textbf{Conciseness}            & \textbf{Expressiveness} \\ 
\midrule
\Ptrnet     & 0.95($\pm$0.10)  &  1.45($\pm$0.18)    & 1.36($\pm$0.04) \\
\Corec      & 0.75($\pm$0.19)  &  1.35($\pm$0.10)    & 1.27($\pm$0.07) \\
\toprule
\multirow{2}{*}{\shortstack{ \revise{Ours}  \\  \revise{(subject)} }} & 1.69($\pm$0.03) & 2.17($\pm$0.40)   & 1.98($\pm$0.19) \\
& \revise{$\uparrow78.17\%$}    & \revise{$\uparrow49.54\%$}    & \revise{$\uparrow45.59\%$}  \\
\midrule
\multirow{2}{*}{\shortstack{ \revise{\bf Ours} }}       & \textbf{2.10($\pm$0.04)}   & \textbf{2.32($\pm$0.40)}   & \textbf{2.16($\pm$0.23)}  \\
& \revise{\textbf{$\uparrow121.83\%$}}   & \revise{\textbf{$\uparrow59.63\%$}}   & \revise{\textbf{$\uparrow58.82\%$}}   \\

\bottomrule
\end{tabular}
\end{table}
The human score results are shown in \Tab~\ref{tab:human-evaluation}. As it shows, the subject part of commit messages generated by our model performs better than \Ptrnet and \Corec consistently. Moreover, providing type and scope information can further improve performance, especially content adequacy than generating only the subject component. 

\subsection{Case Study}
\begin{figure}[h]
\centering
    \includegraphics[width=0.8\columnwidth]{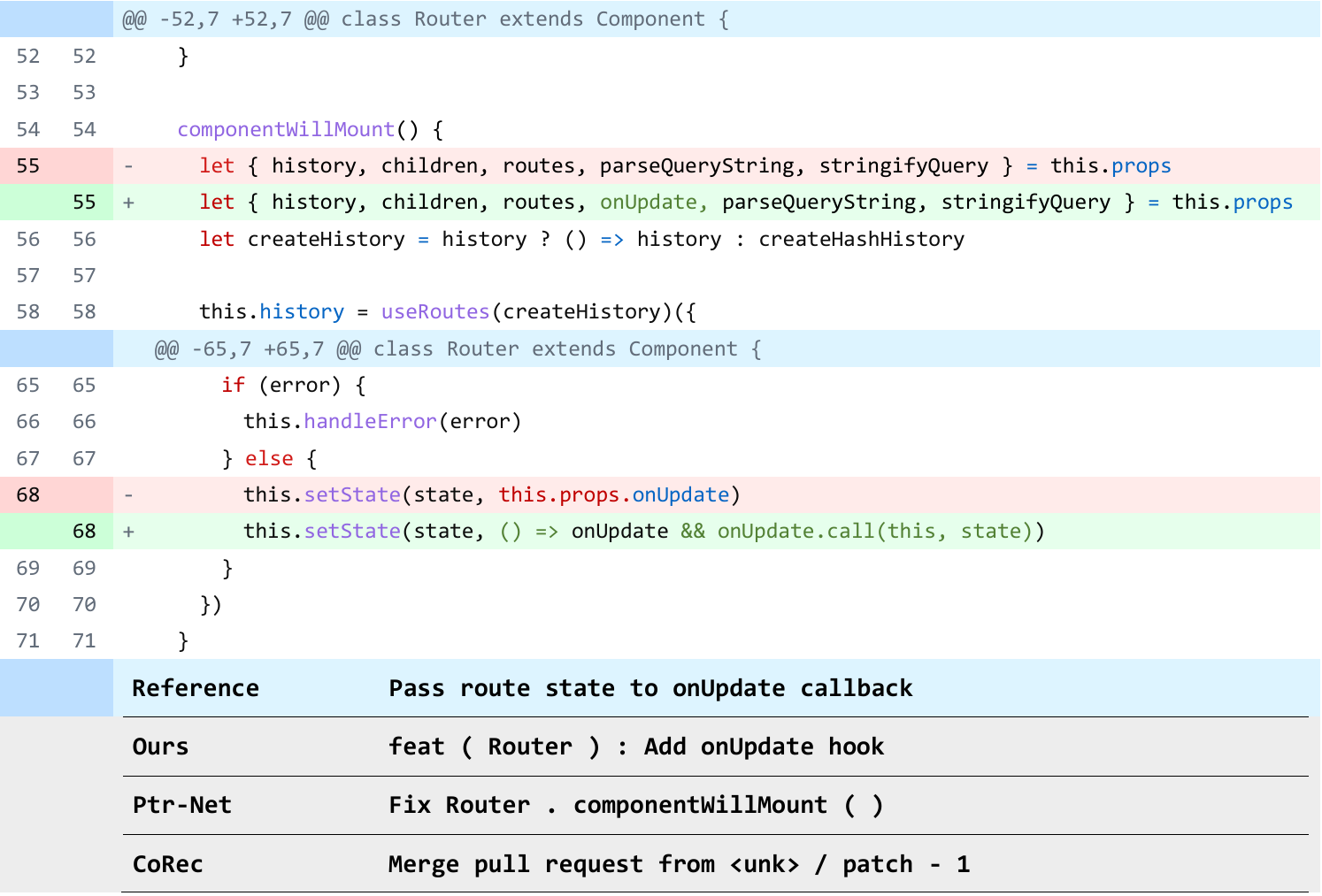}

    \caption{An example of commit and the corresponding commit messages generated by three models.}
    \label{fig:case-study-js}
\end{figure}

\Fig~\ref{fig:case-study-js} shows a commit example\footnote{The figure shows the front part of it, the full part can be found at~\url{https://github.com/remix-run/react-router/commit/3b2ab7e}} from \mcmdjs to compare commit messages generated by different baselines and our models.
The commit message generated by our model provides both why and what code is changed.
From the pull request discussion\footnote{\url{https://github.com/remix-run/react-router/pull/2507}}, we can infer that this commit is aimed to add a new feature, which is denoted as ``feat'' defined by the AngularJS rule\footnote{\url{https://github.com/angular/angular.js/blob/master/DEVELOPERS.md\#commits}}. ``Router'' is the name of a class which is the specifying place of the commit change. Other generated commit messages do not correctly describe what code changes and why these changes are made.
The commit message generated by \Ptrnet also provides the reason for the commit but what changes are made is not described as clearly as our method's generation. \Corec does not provide the correct message for this code change.

\begin{figure}[h]
\centering
    \includegraphics[width=0.7\columnwidth]{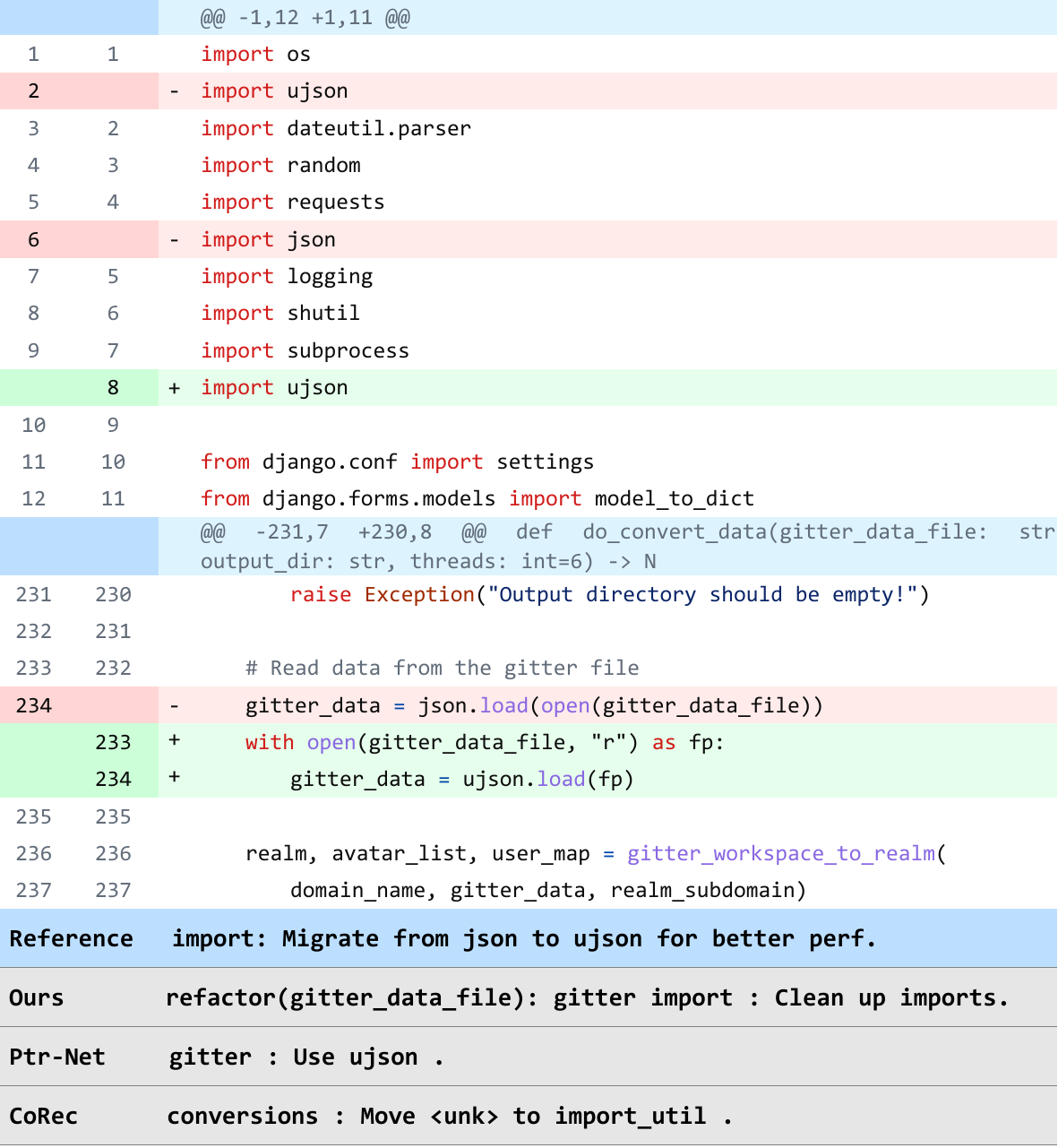}
    \caption{An example of commit and the corresponding commit messages generated by three models.}
    \label{fig:case-study-python}
\end{figure}

Another example\footnote{The figure shows the front part of it, the full part can be found at~\url{https://github.com/zulip/zulip/commit/f9b6eeb}} is shown in \Fig~\ref{fig:case-study-python}. This code change ``neither fixes a bug nor adds a feature'' so this commit belongs to the type ``refactor'', which is similar to ``for better perf'' in reference. Moreover, the scope of this code change is gitter\_data\_file as shown in the line 233-234 of the new-version code. The commit message generated by \Ptrnet and \Corec hardly conveys what code changes are made. The ``type'' and ``scope'' component in our model's generation is helpful to provide the reason for this commit and the ``subject'' component also describe what changes are made in this commit. The commit messages generated by \Ptrnet and \Corec do not provide the correct content.

Above two examples are from \mcmdjs and \mcmdpython. More examples in other PL subsets of \mcmd can be found in our repository\footnote{\url{https://github.com/DeepSoftwareAnalytics/KADEL}}. 

\section{Discussion}

\subsection{Improved Solution for Dataset with Low Rule-Matched Cases}\label{sec:solution4java}

As shown in \Tab~\ref{tab:baselines-comparison-all} and \Tab~\ref{tab:baselines-comparison-all-split-all2sub}, the possible special case in the performance of our model is that the improvement of \meteor score is less than 1\%. As described in \Sec~\ref{sec:perf}, the reason is the low ratio of \mcmdjm in \mcmdjava. To address that issue in the dataset, the knowledge model of our method can also be trained with \mcmdplm in other PL.
In order to control the number of training samples for the knowledge model as in other PL, we try to use \mcmdjsm instead of \mcmdjm for the training of the knowledge model. On the basis of the knowledge model, the subsequent methods remain unchanged. The experimental results in this way are shown in \Tab~\ref{tab:baselines-comparison-java}. These results show that our model can achieve the best performance in \mcmdjava when there are enough rule-matched examples for the knowledge model's training.
Moreover, it also shows that our knowledge model has the potential to be applied in other PLs beyond five PLs of \mcmd.

\begin{table*}[!t]
\centering
\caption{Model performance on the \mcmdjava test set.}
\label{tab:baselines-comparison-java}
\begin{tabular}{llcccccc}
\toprule
\textbf{Dataset} & \bf Metric & \bf \Commitgen    & \bf \Nmt  & \bf \Nngen    & \bf \Ptrnet   & \bf \Corec    & \bf \Ourmethod+    \\
\cmidrule(r){1-2} \cmidrule(r){3-8}
\multirow{3}{*}{\mcmdju}       
& \bf \bleu     & 12.36   & 13.39   & 17.79   & 15.33   & 16.09   & \bf{19.99} $\uparrow11.97\%$\\
& \bf \meteor  & 14.12   & 15.99   & 22.09   & 19.13   & 19.57   & \bf{22.37} $\uparrow1.27\%$\\
& \bf \rouge   & 12.91   & 15.32   & 20.85   & 18.63   & 18.66   & \bf{23.39} $\uparrow12.18\%$\\
\cmidrule(r){1-8}
\multirow{3}{*}{\mcmdjm}       
& \bf \bleu     & 19.77   & 14.01   & 23.03   & 15.17   & 16.47   & \bf{27.16} $\uparrow17.93\%$\\
& \bf \meteor  & 24.29   & 18.40   & 29.85   & 18.91   & 22.02   & \bf{36.88} $\uparrow23.56\%$\\
& \bf \rouge   & 21.72   & 18.31   & 26.70   & 20.13   & 20.14   & \bf{33.73} $\uparrow26.33\%$\\
\toprule
\multirow{3}{*}{\mcmdjava}       
& \bf \bleu     & 12.39   & 13.39   & 17.81   & 15.33   & 16.09   & \bf{20.02} $\uparrow12.41\%$\\
& \bf \meteor  & 14.16   & 16.00   & 22.12   & 19.13   & 19.58   & \bf{22.43} $\uparrow1.40\%$\\
& \bf \rouge   & 12.94   & 15.33   & 20.87   & 18.64   & 18.67   & \bf{23.42} $\uparrow12.22\%$\\
\bottomrule
\end{tabular}
\end{table*}

\subsection{Comparison with ChatGPT}~\label{sec:chatgpt_compar}

ChatGPT~\citep{chatgpt} has attracted attention from both academia and industry since it is announced in November 2022. \revise{Pioneering researchers \citep{geng2023large, from_cmg_history, llm_code_survey_23, DBLP:journals/corr/abs-2308-11396} have effectively utilized ChatGPT, introducing LLM-based methods for generating code comments.}
\revise{To compare our method with ChatGPT, we conduct some simple experiments. The details are described below.}

\revise{Firstly, following the guidelines~\citep{shieh2023best}, we design three types of prompts~\footnote{\revise{full prompt content can be seen in our repository.}} to reduce the impact of prompts: (1) Basic prompt. (2) Basic prompt + Output format. (3) Rephrased prompt by ChatGPT. In the pre-study, these were employed to assess ChatGPT's performance across fifty randomly chosen commits per programming language.

Secondly, after the pre-study, we use a well-designed role prompt template from the study~\citep{abs-2305-12138} with the output format to generate commit messages with ChatGPT~(version: gpt-3.5-turbo-0613)} \revise{on whole test sets in all five programming languages in \mcmd. 
All of the ChatGPT generation results including the prompt can be found in our repository.}

\revise{Thirdly, considering the automatic metrics are focused on the similarity between the reference and the generation rather than the quality of the generated commit message, we also conduct a human evaluation of fifty ChatGPT generation results to evaluate from three perspectives: content adequacy, conciseness, and expressiveness, which can further enhance the solidity of the evaluation. The selected commits in this human evaluation are randomly selected from the test set. They are the same as the commits in the human evaluation described in \Sec~\ref{sec:human-eval} to compare with our model and previous baselines. The standard of the score and the experts who label the data are the same as described in \Sec~\ref{sec:human-eval}.}

\begin{figure}[h]
\centering
    \includegraphics[width=0.66\columnwidth]{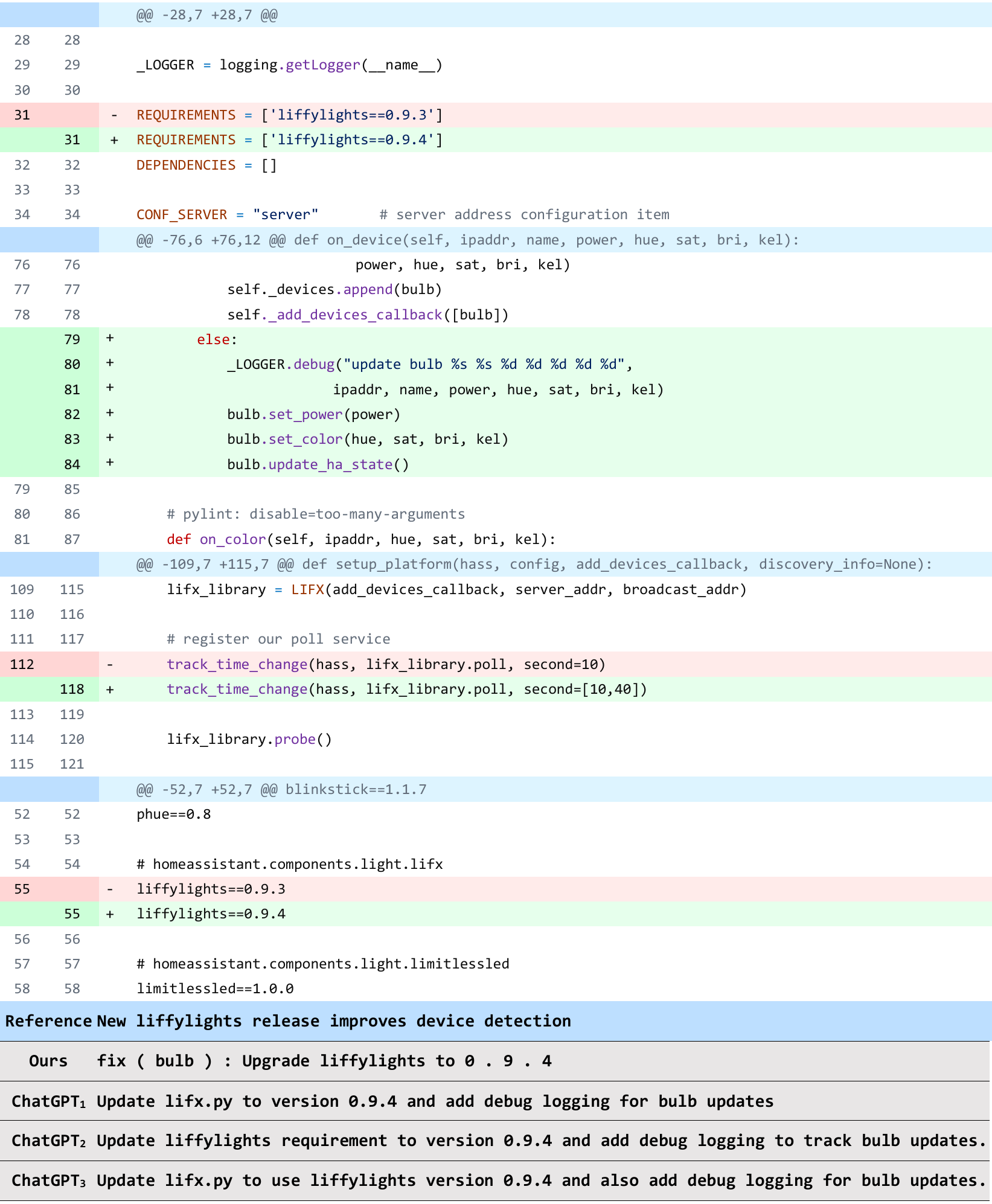}
    \caption{\revise{An example of commit and the corresponding commit messages generated by ChatGPT with three types of prompts. (The subscript indicates the prompt type.)}}
    \label{fig:case-study-chatgpt-1}
\end{figure}

\subsubsection{\revise{Case Study}}
\revise{One example\footnote{\revise{The raw commit can be found at~\url{https://github.com/home-assistant/core/commit/9caa475}}} is shown in \Fig~\ref{fig:case-study-chatgpt-1}. The generated messages by ChatGPT in three types of prompts are similar in many words: the action word ($update$), the version number ($0.9.4$), and the object ($bulb$). All of them are relative to the code change and similar to the reference. It indicates that ChatGPT has a strong ability to deal with code changes to generate commit messages in different types of prompts. Moreover, compared with the reference, the generated result by the second type of prompt is more similar than others because it contains $liffylights$ as the reference. Our model generates similar messages as the $ChatGPT_2$'s generation. As the information in the generation by our model and by $ChatGPT_2$ are nearly the same, we cannot conclude which one is better.}

\begin{figure}[h]
\centering
    \includegraphics[width=0.66\columnwidth]{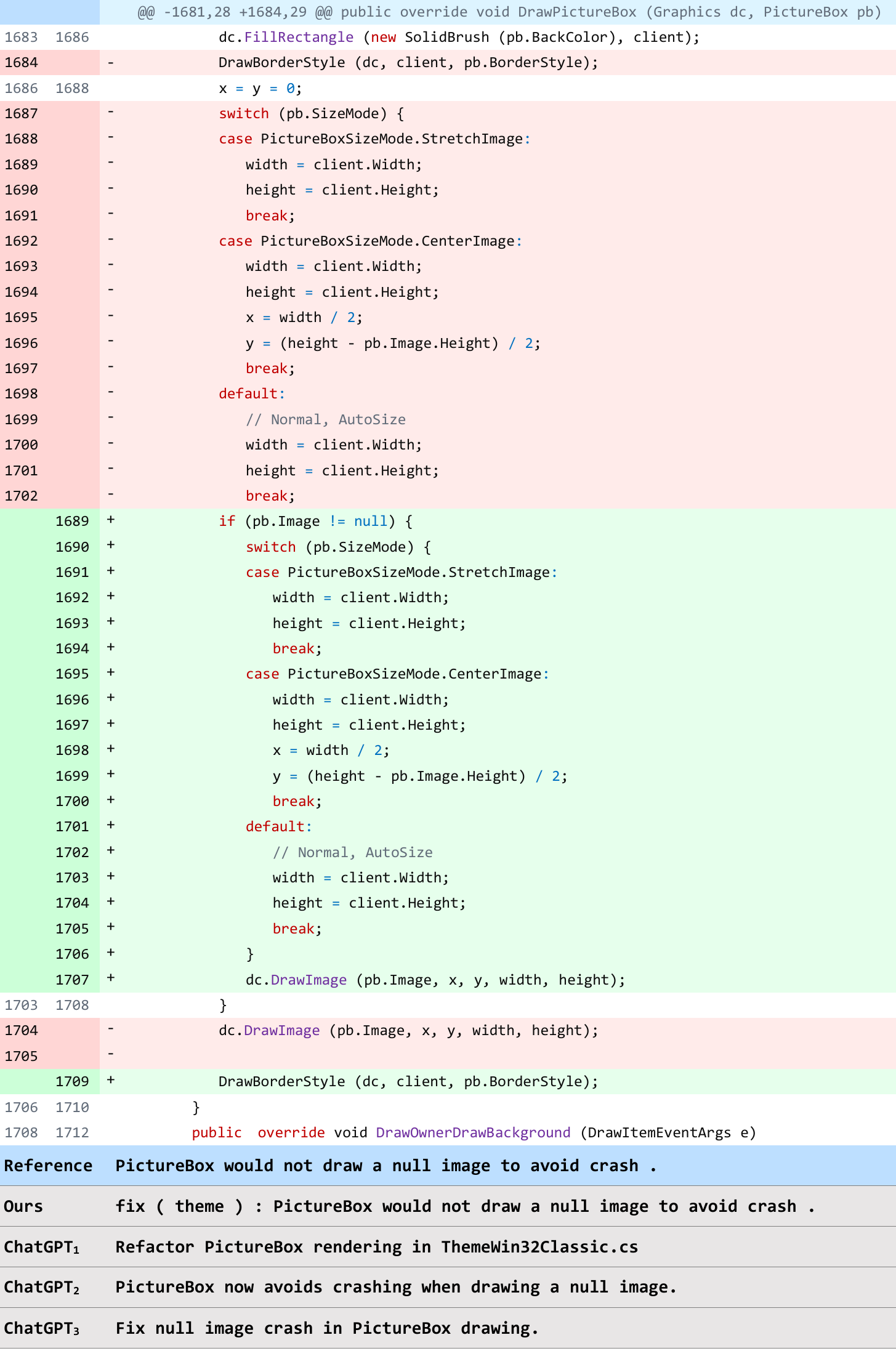}
    \caption{\revise{An example of commit and the corresponding commit messages generated by ChatGPT with three types of prompts. (The subscript indicates the prompt type.)}}
    \label{fig:case-study-chatgpt-2}
\end{figure}

\revise{Another example\footnote{\revise{The figure shows the part of it, the full part can be found at~\url{https://github.com/mono/mono/commit/2ff8c74}}} is shown in \Fig~\ref{fig:case-study-chatgpt-2}. In this case, the second and third prompts show better performance than the first one as they contain many similar words as the reference: $PictureBox$, $null image$, $draw$, and $crash$. These words are key to understanding the commit. Considering the similarity with reference, our model is the best because the ``subject'' part of it is the same as the reference. On the other hand, all of the commit messages generated by our model, $ChatGPT_2$ and $ChatGPT_3$ have similar information and it is difficult to conclude which one is better.}

\begin{figure}[h]
\centering
    \includegraphics[width=0.66\columnwidth]{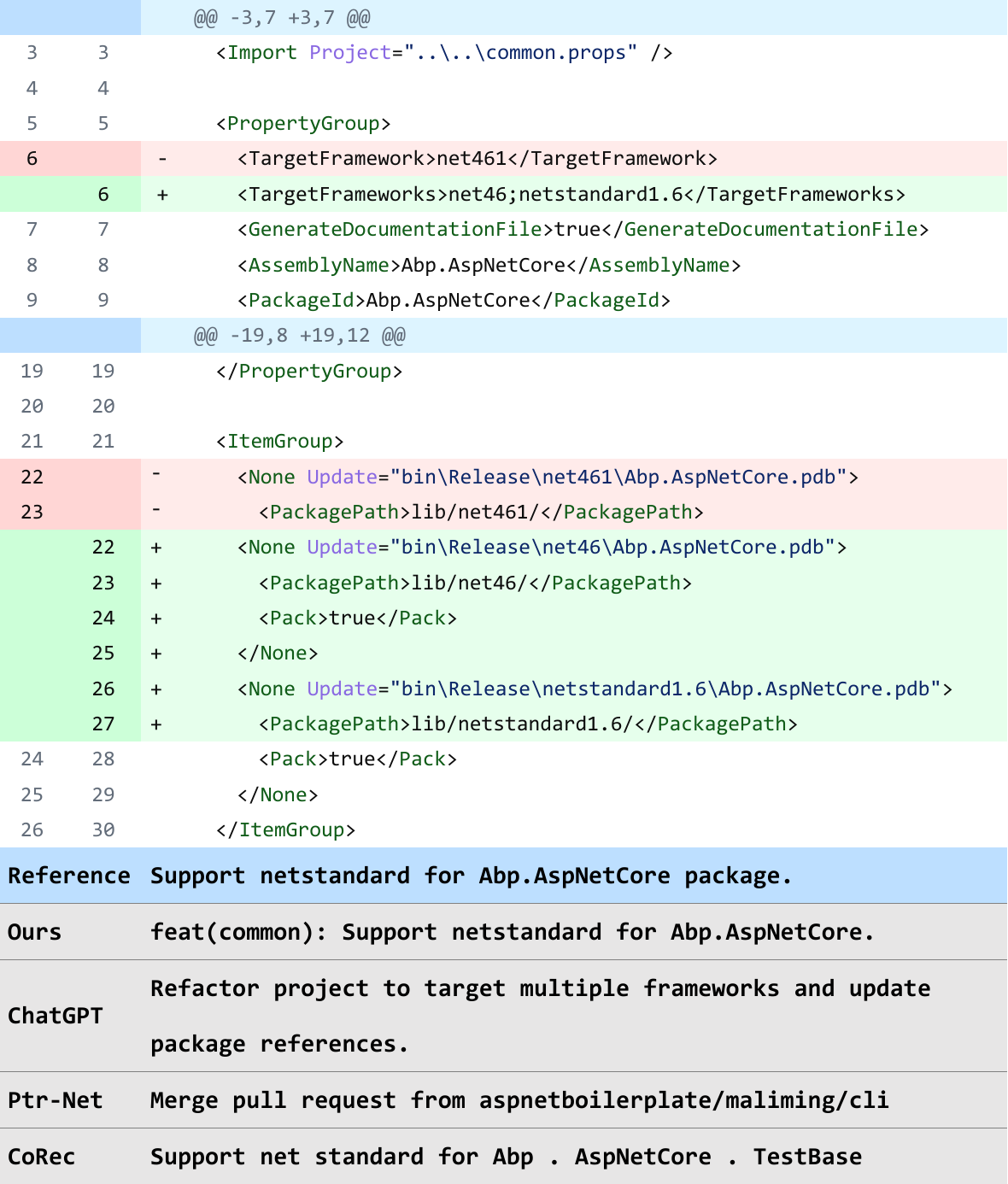}
    \caption{An example of commit and the corresponding commit messages generated by three models and ChatGPT \revise{ with the second prompt}.}
    \label{fig:case-study-csharp}
\end{figure}

\revise{Another} example\footnote{The figure shows the front part of it, the full part can be found at~\url{https://github.com/aspnetboilerplate/aspnetboilerplate/commit/8f548ec}} is shown in \Fig~\ref{fig:case-study-csharp}. Although ChatGPT has a large context window to be inputted with the whole code change, the generation result is not good because it is not consistent with reference (This code change is not refactoring) and the description is not specific (It is not clear what package is updated). In comparison, the generated result of our method is closer to the reference as the subject component of us is the same as the reference. Moreover, the type $feat$ means adding a new feature to support and the scope $common$ is the changed project, which conforms to the content of the code change. This conclusion also holds to the other four PLs' examples.

\subsubsection{\revise{Quantitative Analysis}}
\revise{Automatic evaluation results of ChatGPT are shown in \Tab~\ref{tab:chatgpt-comparison}. As it shows, all the scores of ChatGPT are smaller than the scores of our model among all five programming languages. It indicates that the similarity of the generation between the reference and ChatGPT is less than that between the reference and our model. One possible reason is that ChatGPT is not aimed at one certain domain so it cannot generate similar-style commit messages as reference. Moreover, the difference between BLEU scores is more than that between METEOR. A possible reason is that the ChatGPT is more flexible in expressiveness. The difference in expressiveness also appears in the case study about \Fig~\ref{fig:case-study-chatgpt-2}.}

\revise{Therefore, human evaluation can better evaluate the performance of ChatGPT's generation results. The human evaluation results are shown in Table~\ref{tab:human-evaluation-chatgpt}. This evaluation reveals that ChatGPT's content adequacy is nearly equivalent to that of human (reference), differing by a mere 0.04. ChatGPT outperforms human benchmarks in terms of conciseness and expressiveness. 
Although our model surpasses other baseline models as shown in \Tab~\ref{tab:human-evaluation}, it does not exceed ChatGPT's capabilities. The difference between our model and ChatGPT is less than 0.5 in content adequacy and conciseness, but more pronounced in expressiveness.}

\revise{Overall, the scores of our model range from 2.10 to 2.32, which is higher than the midpoint (2) of the five-point scale (0, 1, 2, 3, 4) so it means the performance is above the moderate level. Although the scores of our model are not higher than the scores of human-written and ChatGPT, all of the scores on content adequacy and conciseness are between the midpoint score (2) and one level above the midpoint score (3). It indicates that the generation performance of our model and ChatGPT is at a similar level.}

\begin{table*}[!t]
\centering
\small
\caption{\revise{ChatGPT performance on each subset of \mcmd test set.}}
\label{tab:chatgpt-comparison}
\begin{tabular}{cllcccccc}
\toprule
\bf \revise{Model} & \bf \revise{Dataset} & \bf \revise{Metrics} 
& \bf \revise{JavaScript} & \bf \revise{C\#} & \bf \revise{Python} & \bf \revise{C++} & \bf \revise{Java} & \bf \revise{\bf Overall}    \\
\cmidrule(r){1-9}

\multirow{9}{*}{{ \revise{ChatGPT} }}& \multirow{3}{*}{\shortstack{ \revise{\mcmdplu} }} 
&  \revise{\bleu} 	&  \revise{10.35} 	&  \revise{8.55} 	&  \revise{10.75} 	&  \revise{8.99} 	&  \revise{9.30} 	&  \revise{9.57} 	\\
&&  \revise{\meteor} 	&  \revise{18.68} 	&  \revise{15.21} 	&  \revise{18.09} 	&  \revise{15.45} 	&  \revise{16.65} 	&  \revise{16.78} 	\\
&&  \revise{\rouge} 	&  \revise{17.37} 	&  \revise{13.32} 	&  \revise{18.29} 	&  \revise{15.32} 	&  \revise{15.43} 	&  \revise{15.92} 	\\
\cmidrule(r){2-9}
& \multirow{3}{*}{\shortstack{ \revise{\mcmdplm} }} 
&  \revise{\bleu} 	&  \revise{12.96} 	&  \revise{10.19} 	&  \revise{11.52} 	&  \revise{10.19} 	&  \revise{10.35} 	&  \revise{12.42} 	\\
&&  \revise{\meteor} 	&  \revise{21.28} 	&  \revise{16.84} 	&  \revise{20.24} 	&  \revise{16.42} 	&  \revise{16.94} 	&  \revise{20.46} 	\\
&&  \revise{\rouge} 	&  \revise{22.68} 	&  \revise{18.78} 	&  \revise{21.66} 	&  \revise{18.77} 	&  \revise{19.01} 	&  \revise{21.98} 	\\
\cmidrule(r){2-9}
& \multirow{3}{*}{\shortstack{ \revise{\mcmd} }} 
&  \revise{\bleu} 	&  \revise{10.58} 	&  \revise{8.56} 	&  \revise{10.75} 	&  \revise{9.00} 	&  \revise{9.30} 	&  \revise{9.64} 	\\
&&  \revise{\meteor} 	&  \revise{18.91} 	&  \revise{15.23} 	&  \revise{18.10} 	&  \revise{15.45} 	&  \revise{16.65} 	&  \revise{16.87} 	\\
&&  \revise{\rouge} 	&  \revise{17.83} 	&  \revise{13.36} 	&  \revise{18.31} 	&  \revise{15.35} 	&  \revise{15.45} 	&  \revise{16.06} 	\\
\cmidrule(r){1-9}

\multirow{9}{*}{{ \revise{KADEL} }}& \multirow{3}{*}{\shortstack{ \revise{\mcmdplu} }} 
&  \revise{\bleu} 	&  \revise{22.86} 	&  \revise{24.72} 	&  \revise{19.99} 	&  \revise{18.21} 	&  \revise{19.79} 	&  \revise{21.08} 	\\
&&  \revise{\meteor} 	&  \revise{26.35} 	&  \revise{26.93} 	&  \revise{23.95} 	&  \revise{20.85} 	&  \revise{22.28} 	&  \revise{24.03} 	\\
&&  \revise{\rouge} 	&  \revise{27.77} 	&  \revise{27.75} 	&  \revise{25.46} 	&  \revise{22.72} 	&  \revise{23.28} 	&  \revise{25.35} 	\\
\cmidrule(r){2-9}
& \multirow{3}{*}{\shortstack{ \revise{\mcmdplm} }} 
&  \revise{\bleu} 	&  \revise{38.28} 	&  \revise{24.95} 	&  \revise{22.61} 	&  \revise{16.94} 	&  \revise{25.27} 	&  \revise{34.55} 	\\
&&  \revise{\meteor} 	&  \revise{51.28} 	&  \revise{36.30} 	&  \revise{39.13} 	&  \revise{31.08} 	&  \revise{34.20} 	&  \revise{47.62} 	\\
&&  \revise{\rouge} 	&  \revise{43.26} 	&  \revise{30.67} 	&  \revise{34.05} 	&  \revise{25.66} 	&  \revise{32.02} 	&  \revise{40.31} 	\\
\cmidrule(r){2-9}
& \multirow{3}{*}{\shortstack{ \revise{\mcmd} }} 
&  \revise{\bleu} 	&  \revise{24.22} 	&  \revise{24.72} 	&  \revise{20.01} 	&  \revise{18.20} 	&  \revise{19.81} 	&  \revise{21.39} 	\\
&&  \revise{\meteor} 	&  \revise{28.55} 	&  \revise{27.00} 	&  \revise{24.07} 	&  \revise{20.93} 	&  \revise{22.33} 	&  \revise{24.58} 	\\
&&  \revise{\rouge} 	&  \revise{29.14} 	&  \revise{27.77} 	&  \revise{25.53} 	&  \revise{22.74} 	&  \revise{23.31} 	&  \revise{25.70} 	\\
\bottomrule

\end{tabular}
\end{table*}

\begin{table}[!t]
\centering 
\caption{\revise{Results of human evaluation (standard deviation in parentheses).}}
\label{tab:human-evaluation-chatgpt}
\begin{tabular}{lccc}
\toprule
\textbf{\revise{Model}}       & \textbf{\revise{Content Adequacy}}        &\textbf{\revise{Conciseness}}            & \textbf{\revise{Expressiveness}} \\ 
\midrule
\revise{ChatGPT}    & \revise{2.55($\pm$0.68)}  & \revise{\bf{2.65($\pm$0.90)}}  & \revise{\bf{3.35($\pm$0.30)}} \\
\revise{Human\footnote{It means the reference commit message written by developers.}}  & \revise{\bf{2.59($\pm$0.16)}}  & \revise{2.51($\pm$0.54)}  & \revise{2.55($\pm$0.54)} \\
\midrule
\revise{Ours}       & \revise{2.10($\pm$0.04)}  & \revise{2.32($\pm$0.40)}  & \revise{2.16($\pm$0.23)}  \\

\bottomrule
\end{tabular}
\end{table}

\subsection{Limitations}

We have identified the following main limitations:
\begin{itemize}
\item{\emph{Human Labeling Bias.}}
The manual annotation of the quality of commit messages may be biased, and inter-rater reliability could be a threat to validity: bias may exist in the scores assigned to the same sentence by different raters. We attempt to mitigate this threat by (1) defining clear scoring rules as shown in \Tab~\ref{tab:human_score_meaning_3} before labeling, and (2) discussing the disagreement cases so that the standard deviations among all raters are reduced.

\item{\emph{Limited Model Scale.}}
In this paper, we employ CodeT5 as our base model rather than large models such as LLaMA~\citep{touvron2023llama}. The reason is two-fold: the effectiveness of CodeT5 and the cost of computational resources.
(1) CodeT5 shows state-of-the-art performance on code-to-text generation tasks in the benchmark CodeXGLUE~\citep{LuGRHSBCDJTLZSZ21}\footnote{\url{https://microsoft.github.io/CodeXGLUE/}}. 
Moreover, as described in \Sec~\ref{sec:chatgpt_compar}, ChatGPT, one of the state-of-the-art large language models, does not show better performance than our method so it is not necessary to incorporate other LLMs in commit message generation.
2) In the tuning setting, other large language models(LLMs) are not selected as the base model to conduct experiments because we do not have huge computing resources. The analysis in this paper requires many experiments, and all of the experiments are conducted on the large dataset, \mcmd, which means that huge computing resources are needed. Training our model based on CodeT5 only needs 5 GPU hours with the support of NVIDIA Tesla V100 (except the time cost of validation and test), which is about 0.02\% of 21K GPU hours that LLaMA consumes in the same setting.

\end{itemize}

\section{Conclusion and Future Work}
In this paper, we empirically find that the commits following the good practice possess an untapped potential and can benefit the pre-trained model to generate better commit messages.
Moreover, we have presented \Ourmethod, a knowledge-aware denoising learning method for the task of commit message generation. Our commit knowledge model is used to learn from data following the good practice. To reduce the negative effects of the noise, we also propose a dynamic denoising training method to learn with commit knowledge more effectively.
Experiments on the large public dataset \mcmd show that compared with previous baselines, \Ourmethod can overall achieve state-of-the-art performance on commit message generation. Our experimental data and code are available at \url{https://github.com/DeepSoftwareAnalytics/KADEL}.

Considering the value of data following the good practice, which incorporates consensus among developers, remains underutilized and not fully explored, the methodology of learning commit knowledge from such data holds significant potential for broader application in diverse tasks. In the future, there is a prospect of extending this methodology to encompass more software engineering tasks.
\revise{Moreover, the comparison with ChatGPT also suggests the potential of LLM for commit message generation. And we will investigate some LLM-based methods for the task.}

\bibliographystyle{ACM-Reference-Format}

\bibliography{main}

\end{document}

%% file: tool.tex
\usepackage{enumitem}
\usepackage{verbatim}
\usepackage{xspace}
\usepackage{diagbox}
\usepackage{array}
\usepackage{cite}

\usepackage{breakurl}
\usepackage{multirow}
\usepackage{diagbox}
\usepackage{url}
\usepackage{comment}
\usepackage{booktabs}
\usepackage{amsmath}
\usepackage{scalerel}
\usepackage[figuresright]{rotating}
\usepackage{threeparttable}

\usepackage{tcolorbox}
\definecolor{grey}{RGB}{128,128,128}

\newcommand{\revise}[1]{{\color{black} #1}}

\newcommand{\mcmd}{MCMD\xspace}
\newcommand{\mcmdjava}{MCMD\textsubscript{Java}\xspace}
\newcommand{\mcmdjm}{MCMD\textsubscript{Java-m}\xspace}
\newcommand{\mcmdju}{MCMD\textsubscript{Java-u}\xspace}
\newcommand{\mcmdcsharp}{MCMD\textsubscript{C\#}\xspace}
\newcommand{\mcmdcsm}{MCMD\textsubscript{C\#-m}\xspace}
\newcommand{\mcmdcsu}{MCMD\textsubscript{C\#-u}\xspace}
\newcommand{\mcmdcpp}{MCMD\textsubscript{C++}\xspace}
\newcommand{\mcmdcppm}{MCMD\textsubscript{C++-m}\xspace}
\newcommand{\mcmdcppu}{MCMD\textsubscript{C++-u}\xspace}
\newcommand{\mcmdpython}{MCMD\textsubscript{Py}\xspace}
\newcommand{\mcmdpym}{MCMD\textsubscript{Py-m}\xspace}
\newcommand{\mcmdpyu}{MCMD\textsubscript{Py-u}\xspace}
\newcommand{\mcmdjs}{MCMD\textsubscript{JS}\xspace}
\newcommand{\mcmdjsm}{MCMD\textsubscript{JS-m}\xspace}
\newcommand{\mcmdjsu}{MCMD\textsubscript{JS-u}\xspace}
\newcommand{\mcmdm}{MCMD\textsubscript{all-m}\xspace}
\newcommand{\mcmdu}{MCMD\textsubscript{all-u}\xspace}
\newcommand{\mcmdpl}{MCMD\textsubscript{PL}\xspace}
\newcommand{\mcmdplm}{MCMD\textsubscript{PL-m}\xspace}
\newcommand{\mcmdplu}{MCMD\textsubscript{PL-u}\xspace}

\newcommand{\Ourmethod}{KADEL\xspace}
\newcommand{\Race}{RACE\xspace}
\newcommand{\Fira}{FIRA\xspace}

\newcommand{\Commitgen}{CmtGen\xspace}
\newcommand{\Nmt}{NMT\xspace}

\newcommand{\Nngen}{NNGen\xspace}

\newcommand{\Ptrnet}{Ptr-Net\xspace} 

\newcommand{\Atom}{ATOM\xspace}

\newcommand{\Codetfive}{CodeT5\xspace}

\newcommand{\Corec}{CoRec\xspace}

\newcommand{\cdiff}{Code Change}

\newcommand{\type}{type\xspace}
\newcommand{\scope}{scope\xspace}
\newcommand{\subject}{subject\xspace}

\newcommand{\bleu}{BLEU\xspace}
\newcommand{\bleunorm}{BLEU-Norm\xspace}

\newcommand{\bleumosesnorm}{BLEU-Norm\xspace}

\newcommand{\meteor}{METEOR\xspace}
\newcommand{\rouge}{ROUGE\xspace}

\newcommand{\rougel}{\rouge-L\xspace}

\newcommand{\Fig}{Figure\xspace}
\newcommand{\Tab}{Table\xspace}
\newcommand{\Sec}{Section\xspace}

\usepackage{mathtools}
\usepackage{multirow}
\usepackage{multicol}
\usepackage{booktabs}
\usepackage{subfigure}

\DeclareMathOperator*{\transformerenc}{Trans-Enc}
\DeclareMathOperator*{\transformerdec}{Trans-Dec}
\DeclareMathOperator*{\transformer}{Transformer}

\DeclareMathOperator*{\emal}{EM}

\usepackage{amsmath,amsfonts,bm}

\def\eqref#1{equation~\ref{#1}}

\def\1{\bm{1}}

\def\rs{{\textnormal{s}}}
\def\rt{{\textnormal{t}}}

\def\vp{{\bm{p}}}

\def\mH{{\bm{H}}}

\def\mL{{\bm{L}}}

\DeclareMathAlphabet{\mathsfit}{\encodingdefault}{\sfdefault}{m}{sl}
\SetMathAlphabet{\mathsfit}{bold}{\encodingdefault}{\sfdefault}{bx}{n}

\def\gC{{\mathcal{C}}}
\def\gD{{\mathcal{D}}}

\def\gL{{\mathcal{L}}}

\def\gV{{\mathcal{V}}}

